\documentclass[a4paper, accepted=2025-06-16, onecolumn]{quantumarticle}
\pdfoutput=1

\usepackage[numbers,sort&compress]{natbib}  

\usepackage[colorlinks = true, linkcolor = quantumviolet, urlcolor  = quantumviolet, citecolor = quantumviolet]{hyperref}
\usepackage{nameref}
\RequirePackage{nameref}
\usepackage{algorithm}
\usepackage{algorithmic}

\usepackage{thmtools, thm-restate}
\usepackage{amsmath}
\usepackage{amsfonts}
\usepackage{amssymb}
\usepackage{amsthm}
\usepackage{bm}
\usepackage{bbold}
\usepackage{empheq}
\usepackage{mathtools}
\usepackage{esvect}

\usepackage{titlesec}
\usepackage{float}

\usepackage{caption}
\usepackage{subcaption}
\usepackage{multirow}

\usepackage{xcolor}

\usepackage{array}

\DeclareMathOperator{\PRO}{PRO}
\DeclareMathOperator{\GME}{GME}

\begin{document}

\title{Estimating the entanglement of random multipartite quantum states}

\author{Khurshed P. Fitter}
\affiliation{École Polytechnique Fédérale de Lausanne, Switzerland}
\email{khurshed.fitter@epfl.ch}

\author{Cécilia Lancien}
\affiliation{Institut Fourier, Université Grenoble Alpes, CNRS, France}
\email{cecilia.lancien@univ-grenoble-alpes.fr}

\author{Ion Nechita}
\affiliation{Laboratoire de Physique Théorique, Université de Toulouse, UPS, CNRS, France}
\email{ion.nechita@univ-tlse3.fr}

\maketitle

\begin{abstract}
Genuine multipartite entanglement of a given multipartite pure quantum state can be quantified through its geometric measure of entanglement, which, up to logarithms, is simply the maximum overlap of the corresponding unit tensor with product unit tensors, a quantity that is also known as the injective norm of the tensor. Our general goal in this work is to estimate this injective norm of randomly sampled tensors. To this end, we study and compare various algorithms, based either on the widely used alternating least squares method or on a novel normalized gradient descent approach, and suited to either symmetrized or non-symmetrized random tensors. We first benchmark their respective performances on the case of symmetrized real Gaussian tensors, whose asymptotic average injective norm is known analytically. Having established that our proposed normalized gradient descent algorithm generally performs best, we then use it to obtain numerical estimates for the average injective norm of complex Gaussian tensors (i.e.,~up to normalization, uniformly distributed multipartite pure quantum states), with or without permutation-invariance. We also estimate the average injective norm of random matrix product states constructed from Gaussian local tensors, with or without translation-invariance. All these results constitute the first numerical estimates on the amount of genuinely multipartite entanglement typically present in various models of random multipartite pure states. Finally, motivated by our numerical results, we posit two conjectures on the injective norms of random Gaussian tensors (real and complex) and Gaussian MPS in the asymptotic limit of the physical dimension.
\end{abstract}

{
  \hypersetup{linkcolor=quantumviolet}
  \tableofcontents
}
\vspace{4mm}


\section{Introduction}
In quantum information theory, the notion of \emph{entanglement} \cite{horodecki2009quantum} plays a central role, providing the framework in which purely quantum correlations are described. Bipartite entanglement, i.e.,~the quantum correlations between two parties, has received a lot of attention in the literature, with a rather simple characterization in the case of pure states, in terms of the \emph{entropy of entanglement}. The case of multipartite entanglement, i.e.,~quantum correlations shared by three or more parties, is qualitatively more complicated. The geometry of the set of pure entangled states is much richer in this case \cite{dur2000three}, and there are several inequivalent ways to measure how much entanglement is present in a given multipartite pure quantum state \cite{coffman2000distributed,barnum2001monotones,wong2001potential}.\\

Among the measures of multipartite pure state entanglement discussed in the literature, the \emph{geometric measure of entanglement} (GME) introduced in \cite{barnum2001monotones} has a natural and geometrical definition: it is the (negative logarithm of the) maximum overlap between the given quantum state and any product state. From a mathematical standpoint, this quantity is identical to the \emph{injective tensor norm} of the corresponding tensor \cite{grothendieck1956resume,ryan2013introduction,aubrun2017alice}. The latter quantity is the focus of this work: we discuss several numerical algorithms to compute the injective norm, particularly for various models of random multipartite pure states.\\

Random quantum states have found several applications in quantum information theory \cite{collins2016random}, ranging from the characterization of entanglement of mixed bipartite states \cite{aubrun2014entanglement} to the solutions of important conjectures in quantum Shannon theory \cite{hastings2009superadditivity}. In this work, we are concerned with the entanglement of random pure states. The uniform probability distribution on the set of pure states corresponds to the uniform probability distribution on the unit sphere of a complex Hilbert space \cite{zyczkowski2011generating}. Here, we are interested in the case where the total Hilbert space admits a tensor product structure with three or more subspaces. Understanding the amount of entanglement (via the geometric measure) for such random pure states tells us what is the \emph{entanglement of a typical multipartite pure quantum state}, and it is this random variable that we aim to study in this work. We are also interested in estimating the entanglement of random \emph{matrix product states }(MPS), again using a natural probability measure on the set of MPS with given local dimensions \cite{garnerone2010typicality,collins2013matrix,gonzalezguillen2018spectral,lancien2022correlation,haferkamp2021emergent}. \\

The asymptotic regimes that we consider when benchmarking our algorithm correspond to either a large local Hilbert space dimension (few quantum systems with a large number of degrees of freedom) or a large number of subsystems (large number of systems with few degrees of freedom). Both situations are physically motivated, the first one corresponding to large quantum systems (like heat baths) \cite{K_hler_2021}, while the second one being relevant in condensed matter physics \cite{Laflorencie_2016, Bianchi_2022, Hamma_2012}. Understanding the entanglement of typical pure quantum states in these regimes, and especially the way the entanglement (as measured by the GME) scales with the growing parameter, is one of the main motivations of the numerical methods we develop here.\\

As is the case with most questions related to tensors, computing the geometric measure of entanglement, or equivalently the injective norm, is a computationally difficult (NP-hard) question \cite{hillar2013most}. We tackle this problem with various numerical algorithms, obtaining lower bounds on the injective norm by finding good candidates for product states having large overlaps with the input tensor. We compare the widely used alternating least squares algorithm \cite{takane_1977} with our proposed normalized gradient descent algorithm, which incorporates normalization at each step. We benchmark these algorithms on random tensors, with or without symmetry. In the symmetric case, we also test the symmetrized counterparts of these algorithms. We choose symmetrized real Gaussian tensors as the initial benchmark owing to the presence of analytical results for the large-dimensional asymptotic limits of their injective norms \cite{auffinger2013random, spiked}. We notice that the numerical results closely match the analytical values. In the other cases (symmetrized complex tensors and non-symmetrized real and complex tensors), our numerical results for the injective norm of typical tensors provide the first estimates for the almost-sure yet elusive asymptotic values. We observe that the values we obtain match asymptotic lower and upper bounds in the large dimension and/or number of tensor factors limits. We study the effect of symmetrization on the value of the geometric measure of entanglement, showing that, on average, symmetrized tensors are less entangled than their non-symmetrized counterparts. Further, for all the cases, along with the asymptotic values, we predict the leading corrections by fitting function estimators on the numerical values obtained using our normalized gradient descent algorithm.\\

We then present, for the first time, the genuinely multipartite entanglement of random matrix product states (MPS) using the numerical algorithms that we develop. Up to now, the main focus in random MPS literature was the understanding of their bipartite entanglement or correlations, over a bipartition that respects the tensor network structure \cite{garnerone2010typicality,collins2013matrix,gonzalezguillen2018spectral,haferkamp2021emergent,lancien2022correlation}. The novelty of our work is to provide numerical results about the typical multipartite entanglement of such states that are of paramount importance in theoretical physics and condensed matter theory. We compare these values to the ones of Gaussian tensors, showing agreement in the limit of large bond dimension. We also study the effect of introducing translation invariance through (partial) symmetrization on the entanglement of these states. Finally, in addition to our proposed algorithms and numerical results, we posit two conjectures on the asymptotic behavior of the injective norms of random (real and complex) Gaussian tensors and random Gaussian MPS.\\

The paper is organized as follows. In Section \ref{sec:inj_gme}, we introduce the injective norm, its basic properties, and its relation to the quantification of multipartite entanglement via the geometric measure of entanglement. In Section \ref{sec:algos}, we give a detailed exposition of the main numerical algorithms used to estimate the injective norm of arbitrary tensors, and we propose a new algorithm, called normalized gradient descent (NGD), as well as its symmetrized version. Then in Section \ref{sec:benchmark-real}, we benchmark these algorithms on random real Gaussian tensors, with and without permutation symmetry. Having established that our proposed NGD algorithm is the most performant, we then use it, in Section \ref{sec:complex-Gaussian}, to estimate the injective norm of random uniformly distributed multipartite pure quantum states (i.e.,~normalized random complex Gaussian tensors), again with or without permutation symmetry, and formulate our first conjecture. Finally, in Section \ref{sec:MPS}, we study the injective norm of random matrix product states in different scenarios, with or without translation-invariance, and asymptotic regimes (large physical and/or bond dimension), concluding with our second conjecture. In Appendix \ref{appendix:det_states}, we present additional benchmarking results for the different algorithms obtained by running them on deterministic states for which the value of GME is known. In addition, we provide open access to our code containing the proposed algorithms, which can be deployed on any tensor (with any number of tensor factors and any local dimensions) \cite{code}.\looseness-1


\section{The injective norm and the geometric measure of entanglement}\label{sec:inj_gme}
The extent to which a multipartite pure state is entangled can be estimated by the angle it makes with the closest pure product state. The \emph{injective norm} of a tensor \(\Psi \in \mathbb{C}^{d_1}\otimes\mathbb{C}^{d_2}\otimes \cdots \otimes \mathbb{C}^{d_n}\) is defined as

\begin{equation}\label{eqn:1}
    \| \Psi \|_\varepsilon =  \max_{{\hat\phi} \in \PRO} \mid \langle \Psi | \hat{\phi} \rangle \mid,
\end{equation}

where \(\PRO\) is the set of all product states in \(\mathbb{C}^{d_1}\otimes\mathbb{C}^{d_2}\otimes \cdots \otimes \mathbb{C}^{d_n}\), i.e.,~product tensors with Euclidean norm equal to \(1\). Further, the \textit{normalized injective norm} of \(\Psi\) is the ratio of its injective norm to its Euclidean norm and is equal to the injective norm of the normalized tensor \(\hat{\Psi} = \Psi / \|\Psi\|\), where \(\|\cdot\|\) denotes the Euclidean norm.

\begin{equation}
    \|\hat{\Psi}\|_\varepsilon = \frac{\|\Psi\|_\varepsilon}{\|\Psi\|}.
\end{equation}

The \emph{geometric measure of entanglement} (GME) \cite{shimony1995degree,barnum2001monotones,wei2003geometric,zhu2010additivity} is defined as the negative logarithm of \(\| \hat{\Psi} \|_\varepsilon^2\). It is a faithful measure of entanglement for multipartite pure states, i.e.,~it is equal to \(0\) if and only if \(\hat{\Psi}\) is a product state.

\begin{equation}\label{eqn:3}
    \GME(\Psi) = -\log_2\left( \| \hat{\Psi} \|_\varepsilon^2 \right) = -\log_2\left( \frac{\| \Psi \|_\varepsilon^2}{\| \Psi \|^2}\right).
\end{equation}


We also point out that computing the injective norm of a tensor is equivalent to finding its \emph{closest product state approximation}. Consider the objective of finding the closest product state \(\hat{\phi}\in\PRO\) of a tensor \(\Psi \in \mathbb{C}^{d_1}\otimes\mathbb{C}^{d_2}\otimes \cdots \otimes \mathbb{C}^{d_n}\).

\begin{equation}
\begin{split}
    \min_{\hat{\phi} \in \PRO}\|\Psi - \hat{\phi}\|^2   &= \min_{\hat{\phi}\in \PRO} \left(\|\Psi\|^2 + \|\hat{\phi}\|^2 - \langle \Psi | \hat{\phi} \rangle - \langle \hat{\phi} | \Psi \rangle\right)\\ &= \min_{\hat{\phi}\in \PRO} \left(\|\Psi\|^2 + 1 - \langle \Psi | \hat{\phi} \rangle - \langle \hat{\phi} | \Psi \rangle)\right).
\end{split}
\end{equation}

Since \(\|\Psi\|\) is fixed for an initial choice of \(\Psi\), the minimization objective is equivalent to maximizing the inner product between \(\Psi\) and the product state approximation \(\hat{\phi}\). Which in turn, is equivalent to the definition of the injective norm \(\|\Psi\|_\varepsilon\) from Equation \eqref{eqn:1}.

\begin{equation}
\begin{split}
    \min_{\hat{\phi}\in \PRO}\|\Psi - \hat{\phi}\|^2 &= \|\Psi\|^2 + 1  - 2\max_{\hat{\phi}\in \PRO} |\langle \Psi | \hat{\phi} \rangle|\\ &= 1 + \|\Psi\|^2 - 2 \|\Psi\|_\varepsilon.
\end{split}
\end{equation}

We make one last comment about this injective norm. Note that, if \(n=2\), then \(\| \Psi \|_\varepsilon\) is simply the largest Schmidt coefficient of \(\Psi\in \mathbb{C}^{d_1}\otimes\mathbb{C}^{d_2}\) (or equivalently the largest singular value or operator norm, if viewed as an element of \(\mathcal M_{d_1\times d_2}(\mathbb C)\)). It is thus a particularly easy quantity to compute. However, the picture changes drastically for higher order tensors: already for \(n=3\), \(\| \Psi \|_\varepsilon\) is, in general, difficult to estimate (NP-hard), both analytically and numerically \cite{hillar2013most}. \\

An extension of the GME to {mixed quantum states} has been proposed in \cite{barnum2001monotones,wei2003geometric} using the \emph{convex roof construction}. This approach is similar to the one used to define the entanglement of formation, another well-studied entanglement measure \cite{horodecki2009quantum}. Note that in the case of GME, the computational cost of estimating the value for pure states is already high (NP-hard). The extra complexity coming from the convex roof optimization makes it intractable (especially for highly mixed states). For these reasons, we do not discuss the case of mixed states in this work and focus exclusively on pure states.


\section{Algorithms for approximating the injective norm of tensors}\label{sec:algos}
In this section, we briefly explain the algorithms that we employ for finding the closest product state approximation of a tensor.


\begin{figure}[t]
    \centering
    \includegraphics[width=0.95\textwidth]{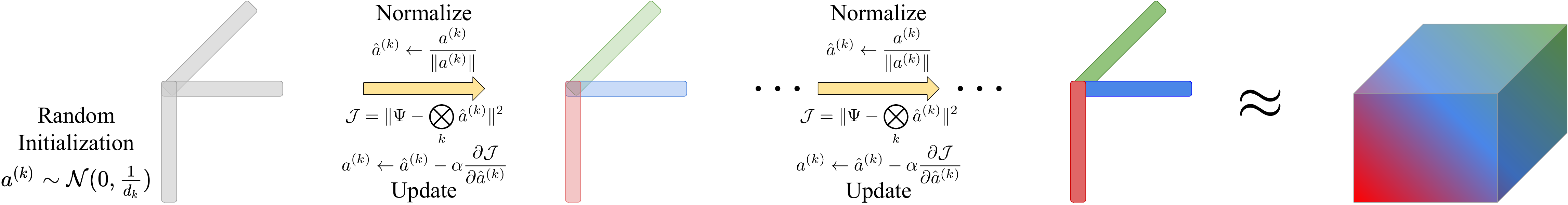}
    \caption{Illustration of our NGD algorithm for finding a rank-1 approximation of a tensor \(\Psi\).}
    \label{fig:ngd}
\end{figure}

\subsection{Alternating least squares}\label{sec:als}
The alternating least squares (ALS) \cite{takane_1977} algorithm tries approximating a tensor \(\Psi\) as a sum of \(R\) rank-1 tensors. The vectors \(a^{(i)}_r \;\forall\; i \in \{1, 2, \ldots, n\}\) involved in calculating the rank-1 tensors are called cores.\looseness-1

\begin{equation}\label{eqn:als}
    \phi = \sum_{r=1}^R a^{(1)}_r \otimes a^{(2)}_r \otimes \cdots  \otimes a^{(n)}_r.
\end{equation}

As the name suggests, ALS focuses on alternatingly optimizing the cores in a way that optimizes the squared error loss between specific views of the approximation and the original tensor \cite{takane_1977}. Since ALS considers only one view of the original tensor at a time, it does not preserve the structure of the tensor during optimization. Further, ALS does not restrict \(\|\phi\|\) during the optimization process, so \(\|\phi\|\) grows in proportion to \(\|\Psi\|\), and it is thus needed to normalize the result post-convergence \(\hat{\phi} = \frac{\phi}{\|\phi\|}\).\looseness-1


\subsection{Power iterations method}\label{sec:pim}
The power iterations method (PIM) \cite{power} is a symmetrized version of ALS that approximates a symmetric tensor as a sum of \(R\) symmetric rank-1 tensors. However, unlike ALS, it incorporates normalization on each step of optimization. Although this eradicates the need for explicitly normalizing the approximation post-convergence, the value of \(\|\phi\|\) at each step of the optimization process still depends on the value of \(\|\Psi\|\). Hence, upon normalizing \(\phi\) on the last step of optimization, PIM induces a similar norm scaling situation as with ALS. 
\begin{equation}\label{eqn:pim}
    \hat{\phi} = \sum_{r=1}^R \hat{a}_r^{\otimes n}.
\end{equation}


\subsection{Normalized gradient descent}\label{sec:ngd}
Although ALS is widely used and is even shown to perform better than gradient descent for finding finite rank approximations of tensors \cite{comon}, it does not consider normalization implicitly. We develop a novel normalized gradient descent (NGD) algorithm as a modification of the projected gradient descent algorithm to approximate tensors as a sum of \(R\) product states. We posit that projectively normalizing the cores on each step of the iterative optimization process restricts the algorithm to optimize over the set of product states. This restriction is crucial since if it is dropped, then the approximated result $\phi$ will require explicit normalization post-convergence. Further, unlike ALS and PIM, the value of \(\|\phi\|\) is completely independent of \(\|\Psi\|\) at all instances. Hence, on every step of the optimization process, the (normalized) approximation is

\begin{equation}\label{eqn:ngd}
    \hat{\phi} = \sum_{r=1}^R \hat{a}^{(1)}_r \otimes \hat{a}^{(2)}_r \otimes \cdots \otimes \hat{a}^{(n)}_r,
\end{equation}

where \(\hat{a}^{(i)}_r\) represents a core \(a^{(i)}_r\) normalized by its Euclidean norm, i.e.,~\(\hat{a}^{(i)}_r = \frac{a^{(i)}_r}{\|a^{(i)}_r\|}\), consequently normalizing $\|\hat\phi\| = 1$. We use the squared error loss as our objective function and calculate the gradients to optimize our objective over each iteration. The squared error loss is simply the square of the Euclidean norm of the difference between the original tensor and its (normalized) approximation.

\begin{equation}\label{eqn:sq_err}
    \mathcal{L}(\Psi, \hat{\phi}) = \|\Psi - \hat{\phi}\|^2.
\end{equation}

Finally, the gradient update step does not preserve normalization, and hence the cores need to be renormalized on each step before computing the approximation.

\begin{algorithm}
\caption{Normalized Gradient Descent to approximate \(\Psi\) with \(\hat\phi \in \PRO\)}
\begin{algorithmic}[1]
    \REQUIRE \(\hat{\phi} \in \PRO \implies \) 
    \ENSURE \(\|\hat{\phi}\|\) = 1\\
    \STATE \textbf{Initialize:} \(a^{(k)}_r \sim \mathcal{N}(0, \frac{1}{d_k}) \;\forall\; r \in 1, 2, \cdots , R, \;\forall\; k \in 1, 2, \cdots , n\)
    \IF{\(\Psi\) is complex}
    \STATE \(a^{\prime(k)}_r \sim \mathcal{N}(0, \frac{1}{d_k}) \;\forall\; r \in 1, 2, \ldots , R, \;\forall\; k \in 1, 2, \ldots , n\)
    \STATE \(a^{(k)}_r \leftarrow (a^{(k)}_r + a^{\prime(k)}_r)/\sqrt{2}\)
    \ENDIF
    \STATE \(t = 0\)
    \FOR{\(t < \texttt{max\_epochs}\)}
    \STATE  \(\hat{a}^{(k)}_r \leftarrow \frac{a^{(k)}_r}{\|a^{(k)}_r\|} \;\forall\; r \in 1, 2, \ldots , R, \;\forall\; k \in 1, 2, \ldots , n\)
    \STATE \(\hat{\phi} = \sum_{r=1}^R \hat{a}^{(1)}_r \otimes \hat{a}^{(2)}_r \otimes \cdots \otimes \hat{a}^{(n)}_r\)
    \STATE \(\mathcal{J}  = \mathcal{L}(\Psi, \hat{\phi})\)
    \STATE \({a}^{(k)}_r \leftarrow \hat{a}^{(k)}_r - \alpha\frac{\partial\mathcal{J}}{\partial\hat{a}^{(k)}_r} \;\forall\; r \in 1, 2, \ldots , R, \;\forall\; k \in 1, 2, \ldots , n\)
    \STATE \(t \leftarrow t + 1\)
    \ENDFOR
\end{algorithmic}
\end{algorithm}


\subsection{Symmetrized gradient descent}\label{sec:sgd}
Analogous to the power iterations method \cite{power}, we develop a symmetrized version of NGD, the symmetrized gradient descent (SGD) algorithm, to approximate symmetric tensors. It optimizes the squared error loss over \(R\) symmetric product states and hence estimates \(R\) cores, with each \(\hat{a}_r \in \mathbb{R}^d\) or \(\mathbb{C}^d \;\forall\; r \in 1, 2, \ldots, R\), leveraging symmetry to reduce the size of the optimization space.
\begin{equation}\label{eqn:sgd}
    \hat{\phi} = \sum_{r=1}^R \hat{a}_r^{\otimes n}.
\end{equation}
For $R=1$, SGD yields an approximation belonging to the symmetrized subspace of product states $\hat\phi \in \operatorname{s{PRO}}$. Similar to NGD, the gradient update step does not preserve normalization, requiring us to renormalize the cores after each update.
\vspace{-2mm}
\begin{algorithm}[H]
\caption{Symmetrized Gradient Descent to approximate symmetrized \(\Psi\) with \(\hat{\phi} \in \operatorname{s{PRO}}\)}
\begin{algorithmic}[1]
    \REQUIRE \(\hat{\phi} \in \operatorname{s{PRO}}\)
    \ENSURE \(\|\hat{\phi}\|\) = 1\\
    \STATE \textbf{Initialize:} \(a_r \sim \mathcal{N}(0, \frac{1}{d_k}) \;\forall\; r \in 1, 2, \cdots , R\)
    \IF{\(\Psi\) is complex}
    \STATE \(a^{\prime(k)}_r \sim \mathcal{N}(0, \frac{1}{d_k}) \;\forall\; r \in 1, 2, \ldots , R, \;\forall\; k \in 1, 2, \ldots , n\)
    \STATE \(a^{(k)}_r \leftarrow (a^{(k)}_r + a^{\prime(k)}_r)/\sqrt{2}\)
    \ENDIF
    \STATE \(t = 0\)
    \FOR{\(t < \texttt{max\_epochs}\)}
    \STATE  \(\hat{a}_r \leftarrow \frac{a_r}{\|a_r\|} \;\forall\; r \in 1, 2, \ldots , R\)
    \STATE \(\hat{\phi} = \sum_{r=1}^R \hat{a}_r^{\otimes n}\)
    \STATE \(\mathcal{J} = \mathcal{L}(\Psi, \hat{\phi})\)
    \STATE \({a}_r \leftarrow \hat{a}_r - \alpha\frac{\partial\mathcal{J}}{\partial\hat{a}_r} \;\forall\; r \in 1, 2, \ldots , R\)
    \STATE \(t \leftarrow t + 1 \)
    \ENDFOR
\end{algorithmic}
\end{algorithm}


\section{Benchmarking the algorithms on random tensors}\label{sec:benchmark-real}

In this section, we test how the algorithms presented in Section \ref{sec:algos} perform. We focus on tensors of order \(2\) and \(3\). The injective norm of order \(2\) tensors is precisely the operator norm of the corresponding matrix, i.e.,~the largest singular value of the operator. This allows us to compare the performance of the algorithms to their established counterparts from linear algebra. The case of order $3$ tensors is fundamentally much more complex \cite{hillar2013most}. Computing the injective norm of order $3$ tensors is already an NP-hard problem \cite[Theorem 1.10]{hillar2013most}. We investigate random tensors with large local dimension $d$, in the spirit of random matrix theory, where most analytical results are obtained in the limit of large matrix dimension. For tensors of order $3$ or more, the only analytical results for the injective norm of random Gaussian tensors are obtained in the asymptotic regime $d \to \infty$, for symmetrized real Gaussian tensors \cite{auffinger2013random,spiked}. Although the algorithms we present work for general tensors of any order, we focus here on order $3$ tensors for two reasons: the computational overheads (time and space complexity) grow exponentially with the order, while order $3$ tensors encompass all the theoretical and computational challenges (NP-hardness) encountered while dealing with higher order tensors. 


\subsection{Random model used for benchmarking}

We estimate the injective norm of a certain class of random tensors in \((\mathbb R^d)^{\otimes n}\), namely real Gaussian tensors and their symmetrized counterparts. Indeed, there are known exact results from previous works for the injective norm of symmetrized real Gaussian tensors \cite{auffinger2013random,spiked}. So we can compare the numerical values obtained by our algorithms to these analytical ones. The choice that we will make for the variance of the entries may seem a bit unintuitive. However, as we will see later, it is a normalization that guarantees that the injective norm of the tensor converges to a finite non-zero value as \(d\) grows.


\subsubsection{Real Gaussian tensors}\label{sec:r_non_gauss}
A real Gaussian tensor \(X \in \left(\mathbb{R}^d\right)^{\otimes n}\) is formed by sampling each of its entries from \(\mathcal{N}\left(0, \frac{2}{d} \right)\). 

\begin{equation}\label{eqn:non_gauss_sample}
X_{i_1, i_2,\ldots , i_n} \sim \mathcal{N}\left( 0, \frac{2}{d} \right) \;\forall\; i_j \in \{1, 2, \ldots , d\}, \;\forall\; j \in \{1, 2, \ldots , n\}.
\end{equation}

The expected value of the squared Euclidean norm of such tensor is simply the dimension of \((\mathbb R^d)^{\otimes n}\), multiplied by the variance factor \(\frac{2}{d}\), i.e.,~\(\frac{2}{d}\times d^n=2d^{n-1}\). The expected Euclidean norm of such a tensor is thus equivalent to \(\sqrt{2d^{n-1}}\) for large \(d\).


\subsubsection{Symmetrized real Gaussian tensors}\label{sec:r_sym_gauss}
A (fully) symmetrized real Gaussian tensor \(X_s \in \left(\mathbb{R}^d\right)^{\otimes n}\) can be formed by projecting a real Gaussian tensor \(X\in \left(\mathbb{R}^d\right)^{\otimes n}\) onto the symmetric subspace of \(\ \left(\mathbb{R}^d\right)^{\otimes n}\). Concretely, we do the following.

\begin{enumerate}
    \item Form an auxiliary Gaussian tensor \(X\in \left(\mathbb{R}^d\right)^{\otimes n}\) with each entry sampled from \(\mathcal{N}\left( 0, \frac{2}{d} \right)\).
    \begin{equation}\label{eqn:sym_gauss_sample}
    X_{i_1, i_2,\ldots , i_n} \sim \mathcal{N}\left( 0, \frac{2}{d} \right) \;\forall\; i_j \in \{1, 2, \ldots , d\}, \;\forall\; j \in \{1, 2, \ldots , n\}.
    \end{equation}
    \item Average over each possible permutation of the auxiliary tensor's axes.
    \begin{equation}
    X_{s_{i_1, i_2,\ldots , i_n}} = \frac{1}{n!}\sum_{\sigma\in S_n}X_{i_{\sigma(1)}, i_{\sigma(2)},\ldots , i_{\sigma(n)}} \;\forall\; i_{\sigma(j)} \in \{1, 2, \ldots , d\}, \;\forall\; j \in \{1, 2, \ldots , n\},
    \end{equation}
    where \(S_n\) is the permutation group of \(n\) elements.
\end{enumerate}

The symmetrized tensor \(X_s\) has its entries sampled from \(\mathcal{N}\big(0, \frac{2}{dn!}\big)\) for locations with no repeated indices. The expected value of the squared Euclidean norm of such a tensor is simply the dimension of the symmetric subspace of \((\mathbb R^d)^{\otimes n}\), multiplied by the variance factor \(\frac{2}{d}\), i.e.,~\(\frac{2}{d} \times \binom{d+n-1}{n} = 2\frac{(d+n-1)\times\cdots\times (d+1)}{n!}\). The expected Euclidean norm of such a tensor is thus equivalent to \(\sqrt{\frac{2}{n!}d^{n-1}}\) for large \(d\).\\


We point out that the factor \(2\) in the variance of the normal distribution in Equations \eqref{eqn:non_gauss_sample} and \eqref{eqn:sym_gauss_sample} is not completely arbitrary. It is indeed the normalization that is standard in the literature regarding the symmetric case \cite{spiked}, since in the case of \(n=2\), where the tensor \(X_s\) can be identified with a symmetric matrix, it guarantees that the off-diagonal terms have variance \(1/d\) while the diagonal ones have variance \(2/d\). We, therefore, use this normalization so that the numerical value we obtain for the injective norm of \(X_s\) can be directly compared to analytical results in the literature, such as those presented in \cite{spiked}. We then use this same normalization in the non-symmetric case just to simplify the analysis: that way the symmetrized tensor has an expected Euclidean norm which is asymptotically (as \(d\rightarrow\infty\)) scaled by a factor \(\frac{1}{\sqrt{n!}}\) as compared to the non-symmetrized tensor.


\subsection{Results on real Gaussian tensors}\label{sec:r_gauss_res}

Analytical results for the injective norms of random tensors are quite limited. It is known that, as \(d\rightarrow\infty\), the injective norm of an order \(n\) symmetric real Gaussian tensor converges to some value \(\eta_n\), which is defined as the unique solution to some explicit equation \cite{auffinger2013random,spiked}. This quantity \(\eta_n\) can thus be estimated numerically for fixed values of \(n\), and its asymptotic behaviour as \(n\rightarrow\infty\) can be identified to be equivalent to \(\sqrt{\log n}\). However, the approach used to obtain this precise asymptotic estimate unfortunately breaks down for non-symmetrized real Gaussian tensors. In this case, only estimates on the order of magnitude of the injective norm can be derived. They tell us that, as \(d\rightarrow\infty\), the injective norm of an order \(n\) (non-symmetrized) real Gaussian tensor converges to some \(n\)-dependent value which scales as \(\sqrt{n\log n}\) for \(n\rightarrow\infty\). A proof that this injective norm is asymptotically \(O(\sqrt{n\log n})\) can be found in \cite{tomioka2014spectral}, while corresponding \(O(\sqrt{n\log n})\) and \(\Omega(\sqrt{n\log n})\) asymptotic estimates are established in the work in progress \cite{lancien2022injective}.\\

\begin{figure}[H]
    \centering
    \includegraphics[width=0.9\textwidth]{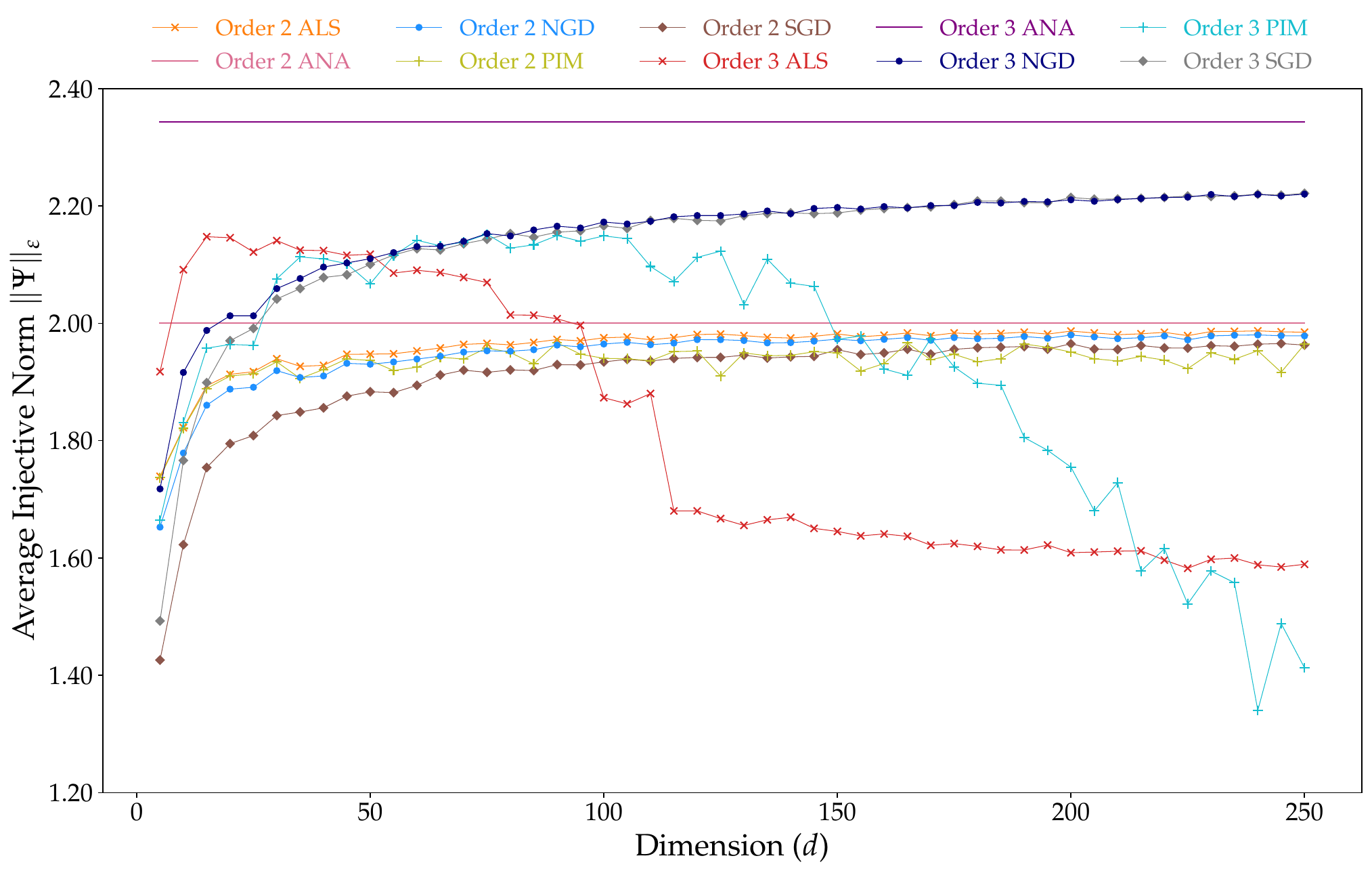}
    \caption{The average injective norm of symmetrized real Gaussian tensors approximated using the ALS, PIM, NGD and SGD algorithms. All the algorithms perform equally well on order \(2\) tensors. NGD and SGD perform substantially better than ALS and PIM on order \(3\) tensors. The analytical bounds are labeled ``ANA''.\looseness-10}\label{fig:r_sym_gauss_inj}
\end{figure}

We first benchmark the performance of all the algorithms on symmetrized real Gaussian tensors because, as explained above, those are the only random tensors for which the asymptotic limit of their injective norm is known. We compare the performance of ALS and NGD along with their symmetrized counterparts, PIM and SGD. For symmetrized Gaussian tensors, the asymptotic values of the injective norm are \(2\) and \(2.343334\) for order \(2\) and order \(3\) respectively, referred from \cite{spiked}. We first estimate the injective norms for tensors with an increasing local dimension $d$, followed by an asymptotic and leading correction estimation using function estimators as described later. As shown in Figure \ref{fig:r_sym_gauss_inj}, for order \(2\) tensors, all algorithms perform equally and converge within \(2.5\%\) of the analytical upper bound. However, for order \(3\) tensors, NGD and SGD show negligible disparity in performance while greatly outperforming ALS and PIM.\\

\begin{figure}[H]
    \centering
    \includegraphics[width=0.9\textwidth]{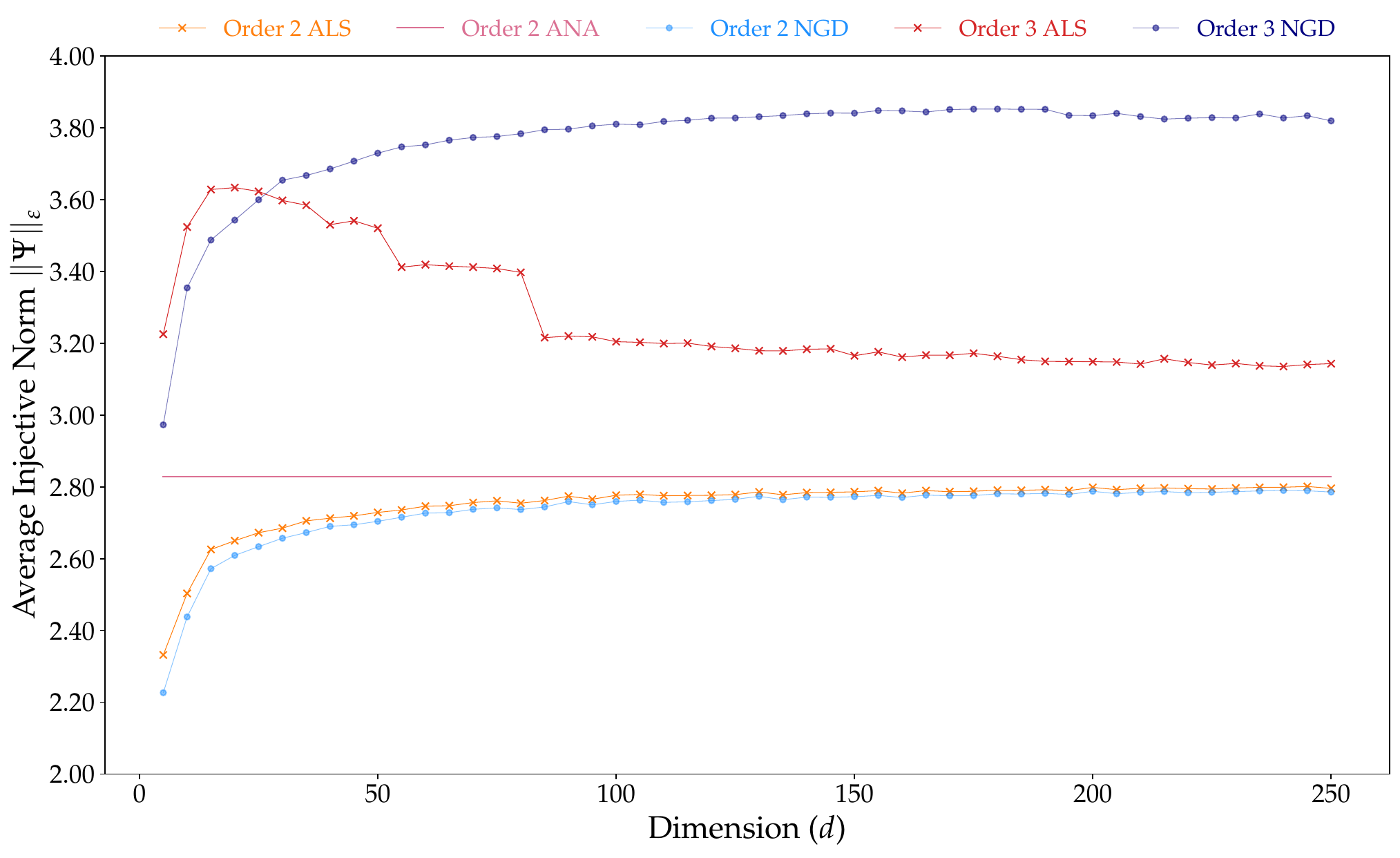}
    \caption{The average injective norm of non-symmetrized real Gaussian tensors approximated using the ALS and NGD algorithms. Both algorithms perform equally well on order \(2\) tensors. NGD performs substantially better than ALS on order \(3\) tensors since both algorithms estimate lower bounds on the unknown analytical value.\looseness-1}
    \label{fig:r_non_gauss_inj}
\end{figure}

Further, we estimate and compare the expected values of the injective norm of non-symmetrized real Gaussian tensors using both ALS and NGD. We know that for order \(2\), the injective norm converges to \(2\sqrt{2}\) as \(d \rightarrow \infty\). Both algorithms perform at par on order \(2\) tensors. Although the analytical value for order \(3\) tensors is unknown, we know that all numerical algorithms estimate a lower bound on the true value of the injective norm. It is thus clear that NGD outperforms ALS by a substantial margin as shown in Figure \ref{fig:r_non_gauss_inj}.\\

We suspect that ALS and PIM perform sub-optimally since the Euclidean norm of the candidate \(\|\phi\|\) depends on that of the original tensor \(\|\Psi\|\). Upon normalizing the ALS result post-convergence, the dependence of \(\|\phi\|\) on \(\|\Psi\|\) induces a sub-optimal lower bound on the injective norm as \(\|\Psi\|\) grows with \(d\). Our proposed NGD and SGD algorithms mitigate this problem by eradicating the dependence of \(\|\phi\|\) on \(\|\Psi\|\) and by restricting \(\phi\) to be a product state $\hat\phi$ through normalization on each step of the iterative optimization process. Further, our proposed NGD algorithm adapts to symmetrization and performs best on both symmetrized and non-symmetrized Gaussian tensors.

\subsection{The effect of normalization}
As mentioned in Sections \ref{sec:als} and \ref{sec:pim}, the norm of the approximation \(\|\phi\|\) is affected by the norm of the original tensor \(\|\Psi\|\) when using the ALS and PIM algorithms. Hence, we also benchmark the performance of all the algorithms on normalized initial tensors \(\hat{\Psi} = \Psi / \|\Psi\|\). In this case, all algorithms perform at par on both symmetrized and non-symmetrized real Gaussian tensors. It can, however, be an issue to be forced to work with normalized inputs. Indeed, in situations such as the ones we are considering here, where the injective norm is much smaller than the Euclidean norm, if the input tensor is normalized, then the output (normalized injective norm) goes to \(0\) as \(d\rightarrow\infty\). To avoid numerical instability and to be able to analyze the results better, it is preferable in such cases to have an algorithm where the input can be scaled so that the output goes to a finite, non-zero, asymptotic value.\\

\begin{figure}[H]
    \centering
    \begin{subfigure}{0.495\textwidth}
    \centering
        \includegraphics[width=\textwidth]{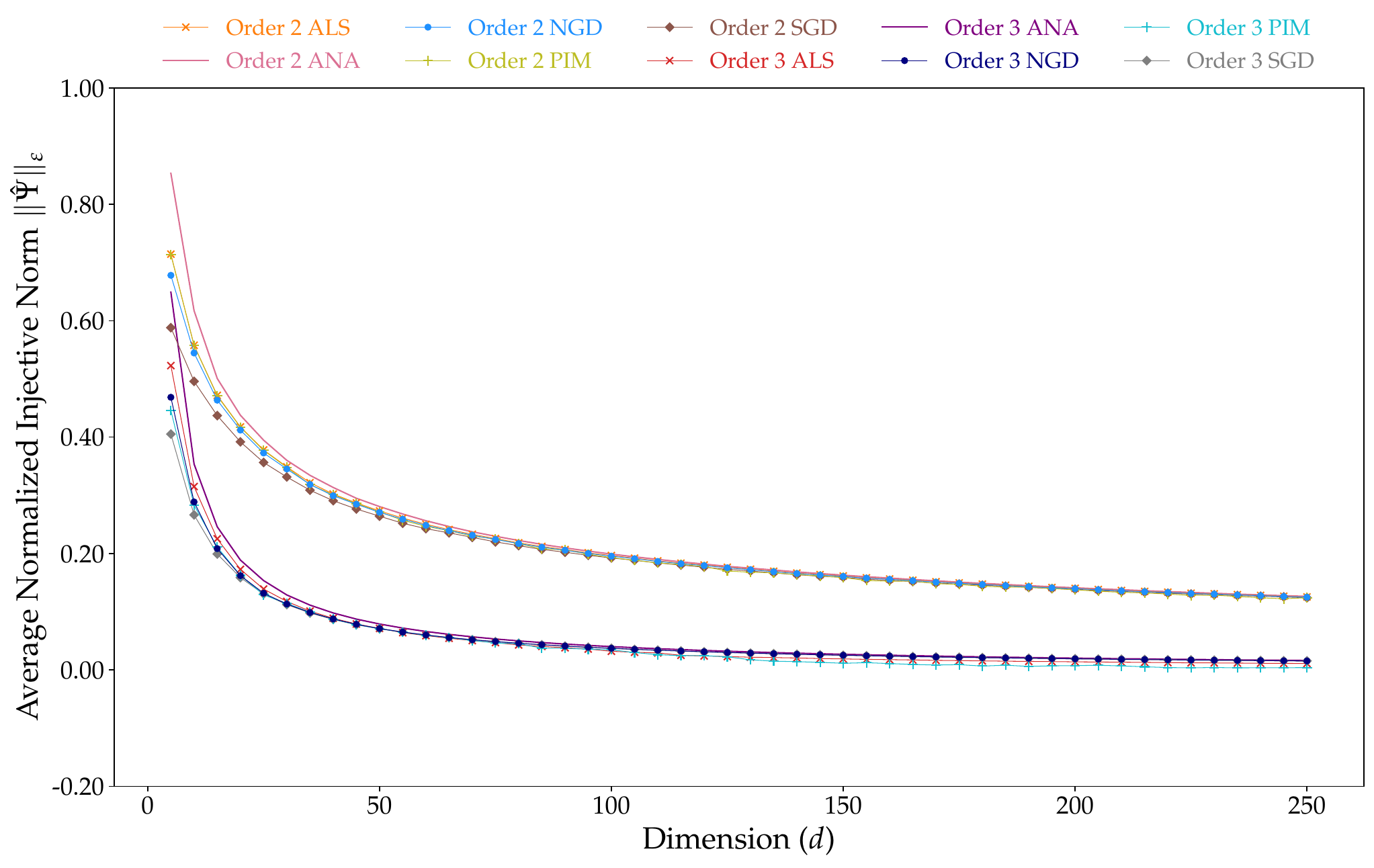}
    \caption{Normalized symmetrized real Gaussian tensors.}
    \label{fig:r_norm_sym_gauss}
    \end{subfigure}
    \begin{subfigure}{0.495\textwidth}
    \centering
        \includegraphics[width=\textwidth]{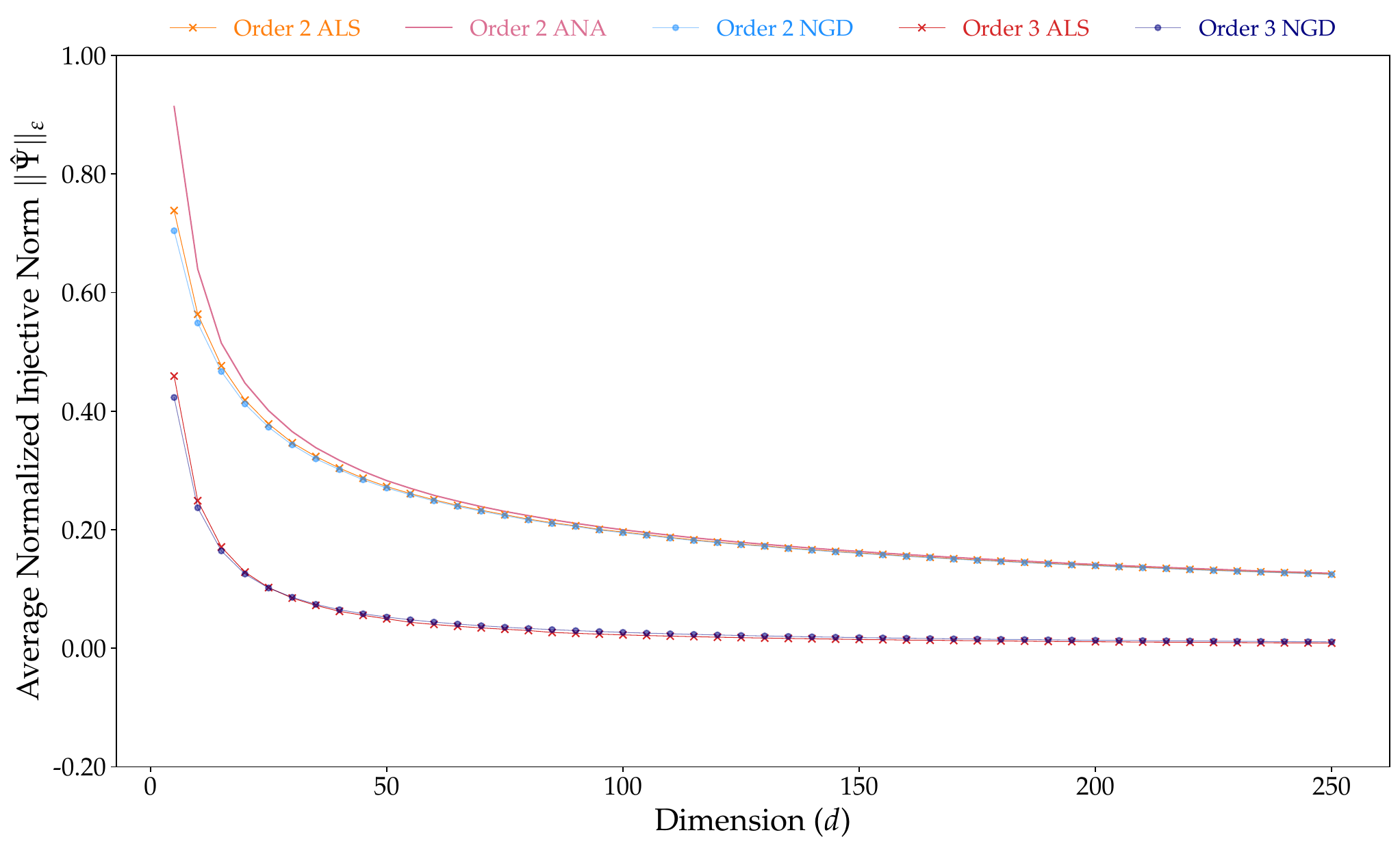}
    \caption{Normalized non-symmetrized real Gaussian tensors.}
    \label{fig:r_norm_non_gauss}
    \end{subfigure}
    \caption{All algorithms perform equally well on inputs that are normalized real Gaussian tensors, either symmetrized or non-symmetrized. However, it is preferable in this case to work with non-normalized tensors because otherwise the injective norm goes to \(0\) as \(d \to \infty\).}
    \label{fig:r_norm_gauss}
\end{figure}

Having established that NGD allows for inputs that are non-normalized (while performing as well as ALS on normalized ones) or non-symmetric (while performing as well as SGD on symmetric ones), it is the algorithm we will systematically use in the sequel. In particular, we will employ it to estimate the injective norm of complex Gaussian tensors (in Section \ref{sec:complex-Gaussian}) and Gaussian matrix product states (in Section \ref{sec:MPS}).\\

On the other hand, in Appendix \ref{appendix:det_states}, we present results that confirm both the ALS and NGD algorithms perform well on normalized tensors. More precisely, we test ALS and NGD on some classes of deterministic multipartite pure states, both sparse and dense, with known values of the GME, and we observe equally good performance from both algorithms. In fact, we point out that we have not found any example of a tensor, either sparse or dense, either deterministic or random, either real or complex, for which our NGD algorithm fails to approximate its injective norm with good accuracy.

\subsection{Numerical function estimators}
As mentioned in Section \ref{sec:r_gauss_res}, the asymptotic value (as \(d\rightarrow\infty\)) of the injective norm is known exactly only for symmetrized real Gaussian tensors, and only its scaling in \(n\) for large $n$ is known for non-symmetrized real Gaussian tensors. Further, even for symmetrized real Gaussian tensors, an exact analytical function for finite \(d\) is still unknown. In this section, we provide numerical estimates of two kinds: (a) for all cases, symmetrized and non-symmetrized, order $2$ and $3$ tensors, we estimate the asymptotes, (b) for the cases where the asymptote is known, i.e.,~symmetrized order $2$ and $3$, non-symmetrized order $2$ tensors, we estimate the leading correction terms.\\

For estimating the asymptotes, we fit a geometric series \((C_0r^k)_{k\in\mathbb N}\) with \(|r| < 1\) on the first differences of our data points. This choice is motivated by the fact that, with the normalization that we choose, as \(d \rightarrow \infty\), the injective norm converges to a fixed constant value and the difference between successive terms vanishes. The infinite sum of the series then gives us the difference between the first data point and the asymptote as \(\frac{C_0}{1-|r|}\). We apply this method to all four cases, order \(2\) and \(3\), symmetrized and non-symmetrized real Gaussian tensors. In the three cases where the analytical value is known, we observe that our obtained numerical estimates match closely with the analytical values \cite{spiked} as shown in Table \ref{table:1}.\\

\vspace{-5mm}
\begin{table}[H]
\centering
\footnotesize{
\begin{tabular}{cllrrrr}
\hline
\multicolumn{1}{c}{Order} &
  Symmetry &
  Field &
\multicolumn{1}{l}{Estimated asymptote \(C_0\)} &
  \multicolumn{1}{l}{Analytical asymptote} &
  \multicolumn{1}{l}{\% Error} \\ \hline
\multirow{2}{*}{2} & \multirow{1}{*}{Symmetrized}     & \(\mathbb{R}\) & \(1.978929\) & \(2.000000\) & \(1.053563\) \\
                   & \multirow{1}{*}{Non-symmetrized} & \(\mathbb{R}\) & \(2.794735\) & \(2.828427\) & \(1.191190\) \\ \hline
\multirow{2}{*}{3} & \multirow{1}{*}{Symmetrized}     & \(\mathbb{R}\) & \(2.248792\) & \(2.343334\) & \(4.034520\) \\
                   & \multirow{1}{*}{Non-symmetrized} & \(\mathbb{R}\) & \(3.949725\) & - & - \\
\hline
\end{tabular}
}
\caption{Estimates for the asymptotic value of the injective norm of real Gaussian tensors.\looseness-1}
\label{table:1}
\end{table}
\vspace{-4mm}
Further, we estimate the leading correction term in the three cases where the asymptote is analytically known. To do this, we fit function estimators of the form \(C_0 + C_1d^\alpha\), where \(C_0\) is a fixed constant (the known analytical asymptote), whereas \(C_1\) and \(\alpha\) are estimated by minimizing the squared error.\\

In the case of order $2$ symmetrized real Gaussian tensors, the theoretical value for $\alpha$ is conjectured to be $-2/3$, due to the following known fact. Let \(G\) be a \(d\times d\) GOE matrix, normalized so that its off-diagonal entries have variance \(\frac{1}{d}\) and its diagonal ones have variance \(\frac{2}{d}\). Note that \(G\) is the same as a symmetric real Gaussian tensor of order \(2\), as defined in Section \ref{sec:r_sym_gauss}, viewed as a matrix rather than a vector. In this matrix picture, the equivalent of the injective norm is simply the largest eigenvalue. We know (see e.g.~\cite[Proposition 6.25]{aubrun2017alice}) that for such random matrix \(G\), for any \(0<\delta<1\),
\[ 
\mathbb P\left(\lambda_{max}(G)>2+\delta\right) \leq e^{-cd\delta^{3/2}} \text{ and } \mathbb P\left(\lambda_{max}(G)<2-\delta\right) \leq e^{-c'd^2\delta^3}. 
\]
Both deviation probabilities are non-vanishing for \(\delta\simeq d^{-2/3}\). So, supposing they are sharp, we expect that, for finite but large \(d\), a good estimator for the behaviour of \(\lambda_{max}(G)\) would be of the form \(2+Cd^{-2/3}\). We observe that the numerical values we obtain for \(\alpha\) in the order \(2\) case, for both symmetrized and non-symmetrized Gaussian tensors, match with this conjectured value of $-2/3$, as shown in Table \ref{table:2}.\looseness-1\\

\begin{figure}[!h]
    \centering
    \begin{subfigure}[b]{0.9\textwidth}
    \centering
    \includegraphics[width=\textwidth]{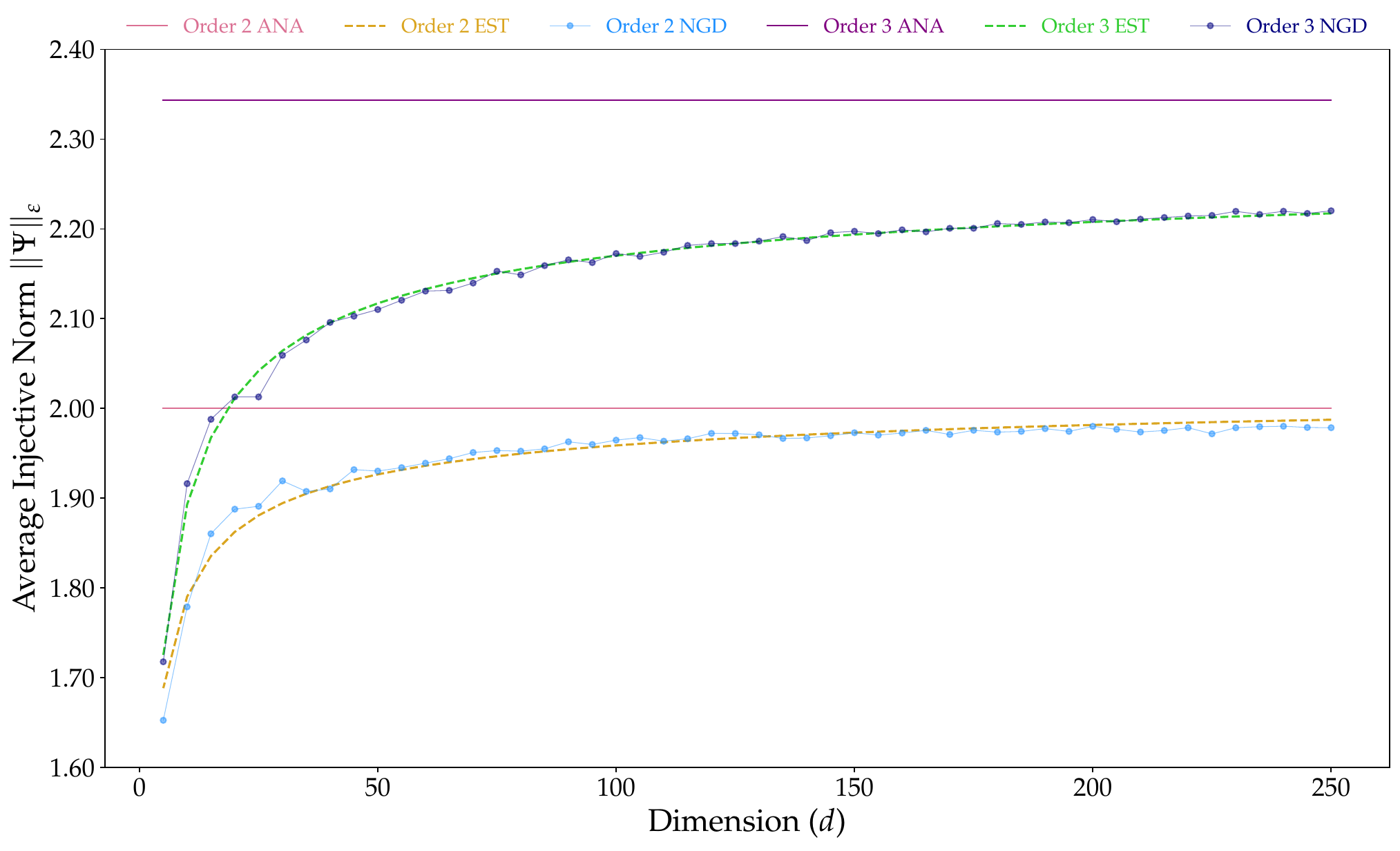}
    \caption{Symmetrized real Gaussian tensors.}
    \label{fig:r_sym_gauss_fit}
    \end{subfigure}

    \centering
    \begin{subfigure}[b]{0.9\textwidth}
    \centering
    \includegraphics[width=\textwidth]{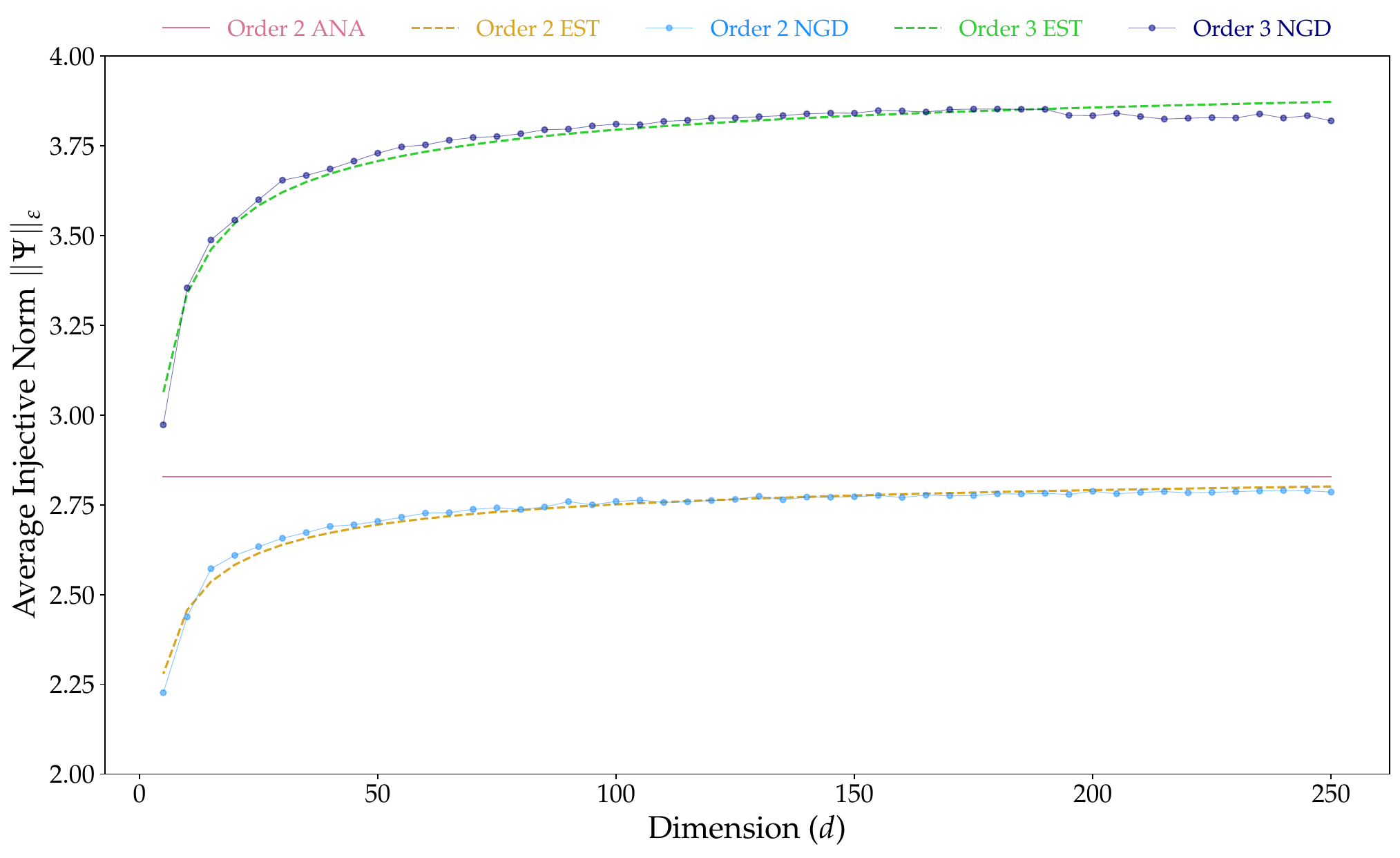}
    \caption{Non-symmetrized real Gaussian tensors.}
    \label{fig:r_non_gauss_fit}
    \end{subfigure}
    \caption{We fit functions of the form \(C_0 + C_1 d^\alpha\) on the data points obtained using our NGD algorithm for symmetrized and non-symmetrized real Gaussian tensors of order \(2\) and \(3\). The asymptotic constant \(C_0\) is fixed to the known analytical value while $C_1$ and $\alpha$ are estimated by minimizing the squared error.}
    \label{fig:r_gauss_fit}
\end{figure}

\begin{table}[H]
\centering
\footnotesize{
\begin{tabular}{cllrrrr}
\hline
\multicolumn{1}{c}{Order} &
  Symmetry &
  Field &
\multicolumn{1}{l}{Analytical asymptote \(C_0\)} &
  \multicolumn{1}{l}{Coefficient \(C_1\)} &
  \multicolumn{1}{l}{Power \(\alpha\)} \\ \hline
\multirow{2}{*}{2} & \multirow{1}{*}{Symmetrized}     & \(\mathbb{R}\) & \(2.000000\) & \(-0.99787802\) & \(-0.69016213\) \\
                   & \multirow{1}{*}{Non-symmetrized} & \(\mathbb{R}\) & \(2.828427\) & \(-1.61362670\) & \(-0.63377984\) \\ \hline
\multirow{2}{*}{3} & \multirow{1}{*}{Symmetrized}     & \(\mathbb{R}\) & \(2.343334\) & \(-1.16629229\) & \(-0.41246088\) \\
                   & \multirow{1}{*}{Non-symmetrized} & \(\mathbb{R}\) & - & - & - \\
\hline
\end{tabular}
}
\caption{Estimates for the leading correction term in the cases where the asymptotic values are analytically known. These values are obtained by fitting function estimators of the form \(C_0 + C_1 d^{\alpha}\), where \(C_0\) is fixed (to the known asymptotic value) while \(C_1\) and \(\alpha\) are estimated by minimizing the squared error. \looseness-1}
\label{table:2}
\end{table}

\section{Complex Gaussian tensors}
\label{sec:complex-Gaussian}

In this section, we look at random complex Gaussian tensors in \(\left(\mathbb{C}^d\right)^{\otimes n}\), starting with a brief description of how we construct these tensors. Having established that our proposed NGD algorithm performs better than the others in all cases, we employ it to obtain lower bounds on the injective norm of complex Gaussian tensors and their symmetrized counterparts. In the complex setting, contrary to the real one, no analytical results are known for the asymptotic value of these norms, even in the symmetric case. Note that a complex Gaussian tensor, after dividing by its Euclidean norm, has the same distribution as a uniform complex unit tensor. Hence, computing the injective norm of a (normalized) complex Gaussian tensor, either symmetrized or not, is equivalent to computing the amount of entanglement in a uniformly distributed multipartite pure state, with or without the constraint of being permutation-invariant. The numerical results for the asymptotes and leading corrections that we obtain are thus the first precise estimates on the amount of entanglement in such random multipartite pure states. They furthermore allow us to quantitatively study the effect of symmetrization on the entanglement present in generic multipartite pure states.


\subsection{Construction}
We construct a complex Gaussian tensor \(X \in \left(\mathbb{C}^d\right)^{\otimes n}\) by sampling the real and imaginary parts of each of its entries from \(\mathcal{N}\big(0, \frac{2}{d}\big)\). Further, we normalize each entry by a factor of \(\sqrt{2}\) to maintain the expected value of the Euclidean norm of the complex Gaussian tensors the same as that of real Gaussian tensors with the same local dimension \(d\) and number of tensor factors \(n\). Similarly, we construct a symmetrized complex Gaussian tensor \(X_s \in \left(\mathbb{C}^d\right)^{\otimes n}\) by first sampling the real and imaginary parts of each entry of the complex Gaussian tensor \(X \in \left(\mathbb{C}^d\right)^{\otimes n}\) from \(\mathcal{N}\big(0, \frac{2}{d}\big)\) and normalizing by a factor of \(\sqrt{2}\) before projecting it onto the symmetric subspace of \(\left(\mathbb{C}^d\right)^{\otimes n}\).


\subsection{Estimating the injective norm}


We estimate the injective norm of complex Gaussian tensors and their symmetrized counterparts using our NGD algorithm. The asymptotic value for order \(2\) tensors is known to be \(2\sqrt{2}\) and \(2\) as \(d \rightarrow \infty\) for the non-symmetrized and symmetrized models, respectively. As mentioned in Section \ref{sec:r_gauss_res}, for higher-order tensors, no exact asymptotic estimate is known. The only thing that can be shown is that, with the normalization that we use for the variance, the injective norm converges to an \(n\)-dependent limit as \(d \rightarrow \infty\), and as \(n \rightarrow \infty\) the limit scales as \(\sqrt{n\log n}\) in the non-symmetrized case and as \(\sqrt{\log n}\) in the symmetrized case. This asymptotic scaling is exactly the same as for real Gaussian tensors in both the symmetrized and the non-symmetrized cases. It is explained how to derive such estimates on the order of magnitude of these norms in \cite{aubrun2017alice}. The precise computations are done in the work in progress \cite{lancien2022injective}. We emphasize that it is not clear at all whether the techniques used in \cite{auffinger2013random} to get the exact asymptotic estimates for the injective norm of symmetrized real Gaussian tensors could be applied to treat even the case of symmetrized complex Gaussian tensors.\\

The numerical results that we obtain for the injective norm of order \(2\) and order \(3\) complex Gaussian tensors are shown in Figures \ref{fig:c_non_gauss_inj} (non-symmetrized case) and \ref{fig:c_sym_gauss_inj} (symmetrized case). In the complex case, we estimate the asymptotic value and the leading correction term using the same two-step strategy as in the real case, with function estimators of the same form.\\

We highlight that, for order $2$ Gaussian tensors, the analytical asymptotic values are the same for real and complex symmetrized tensors on the one hand, and for real and complex non-symmetrized tensors on the other hand. Numerically, we observe that this seems to be the case also for order $3$ Gaussian tensors. All numerical estimates for the asymptotic values are presented in Table \ref{table:3}. Motivated by this evidence, we conjecture the following.
\newtheorem{conj}{Conjecture}
\begin{conj}
 The asymptotic limits of the injective norms for corresponding (same order and symmetry) real and complex Gaussian tensors are identical.
\end{conj} 

Further, we estimate the leading correction term by fitting function estimators of the form \(C_0 + C_1 d^\alpha\), like in the real case. Building upon our conjecture, we use the same value for \(C_0\) in the order \(3\) symmetrized complex Gaussian case as in the known real case, and observe that the correction coefficient and power match closely with those in the real case, as shown in Tables \ref{table:2} and \ref{table:4}.\looseness-10\\

\vspace{-2.5mm}
\begin{table}[H]
\centering
\footnotesize{
\begin{tabular}{cllrrrr}
\hline
\multicolumn{1}{c}{Order} &
  Symmetry &
  Field &
\multicolumn{1}{l}{Estimated asymptote \(C_0\)} &
  \multicolumn{1}{l}{Analytical asymptote} &
  \multicolumn{1}{l}{\% Error} \\ \hline
\multirow{2}{*}{2} & \multirow{1}{*}{Symmetrized}     & \(\mathbb{C}\) & \(2.092316\) & \(2.000000\) & \(4.6158\) \\
                   & \multirow{1}{*}{Non-symmetrized} & \(\mathbb{C}\) & \(2.946784\) & \(2.828427\) & \(4.1846\) \\ \hline
\multirow{2}{*}{3} & \multirow{1}{*}{Symmetrized}     & \(\mathbb{C}\) & \(2.356248\) & \(-\) & \(-\) \\
                   & \multirow{1}{*}{Non-symmetrized} & \(\mathbb{C}\) & \(4.143529\) & \(-\) & \(-\) \\
\hline
\end{tabular}
}
\vspace{-2mm}
\caption{Estimates for the asymptotes for complex Gaussian tensors. Unlike the real case, in the complex case, we do not know the analytical value for the order 3 symmetrized Gaussian tensors. However, we observe that the numerical value is quite close to the real case, and hence we conjecture that they are indeed the same value.}

\label{table:3}
\end{table}

\vspace{-4mm}
\begin{table}[H]
\centering
\footnotesize{
\begin{tabular}{cllrrrr}
\hline
\multicolumn{1}{c}{Order} &
  Symmetry &
  Field &
\multicolumn{1}{l}{Analytical asymptote \(C_0\)} &
  \multicolumn{1}{l}{Coefficient \(C_1\)} &
  \multicolumn{1}{l}{Power \(\alpha\)} \\ \hline
\multirow{2}{*}{2} & \multirow{1}{*}{Symmetrized}     & \(\mathbb{C}\) & \(2.000000\) & \(-0.83908193\) & \(-0.63376019\) \\
                   & \multirow{1}{*}{Non-symmetrized} & \(\mathbb{C}\) & \(2.828427\) & \(-1.68679358\) & \(-0.63961335\) \\ \hline
\multirow{2}{*}{3} & \multirow{1}{*}{Symmetrized}     & \(\mathbb{C}\) & \(2.343334\) & \(-0.90616281\) & \(-0.39636923\) \\
 
                   & \multirow{1}{*}{Non-symmetrized} & \(\mathbb{C}\) & - & - & - \\
\hline
\end{tabular}
}
\vspace{-2mm}
\caption{Estimates for the leading correction term in the cases where the asymptotic values are analytically known or conjectured. The leading correction coefficient and power are obtained by fitting curves of the form \(C_0 + C_1 d^{\alpha}\), where \(C_0\) is fixed (to the known or conjectured asymptotic value), while \(C_1\) and \(\alpha\) are estimated by minimizing the squared error. \looseness-1}
\label{table:4}
\end{table}

\vspace{-5mm}

\begin{figure}[H]
    \centering
    \begin{subfigure}[b]{0.9\textwidth}
     \centering
     \includegraphics[width=\textwidth]{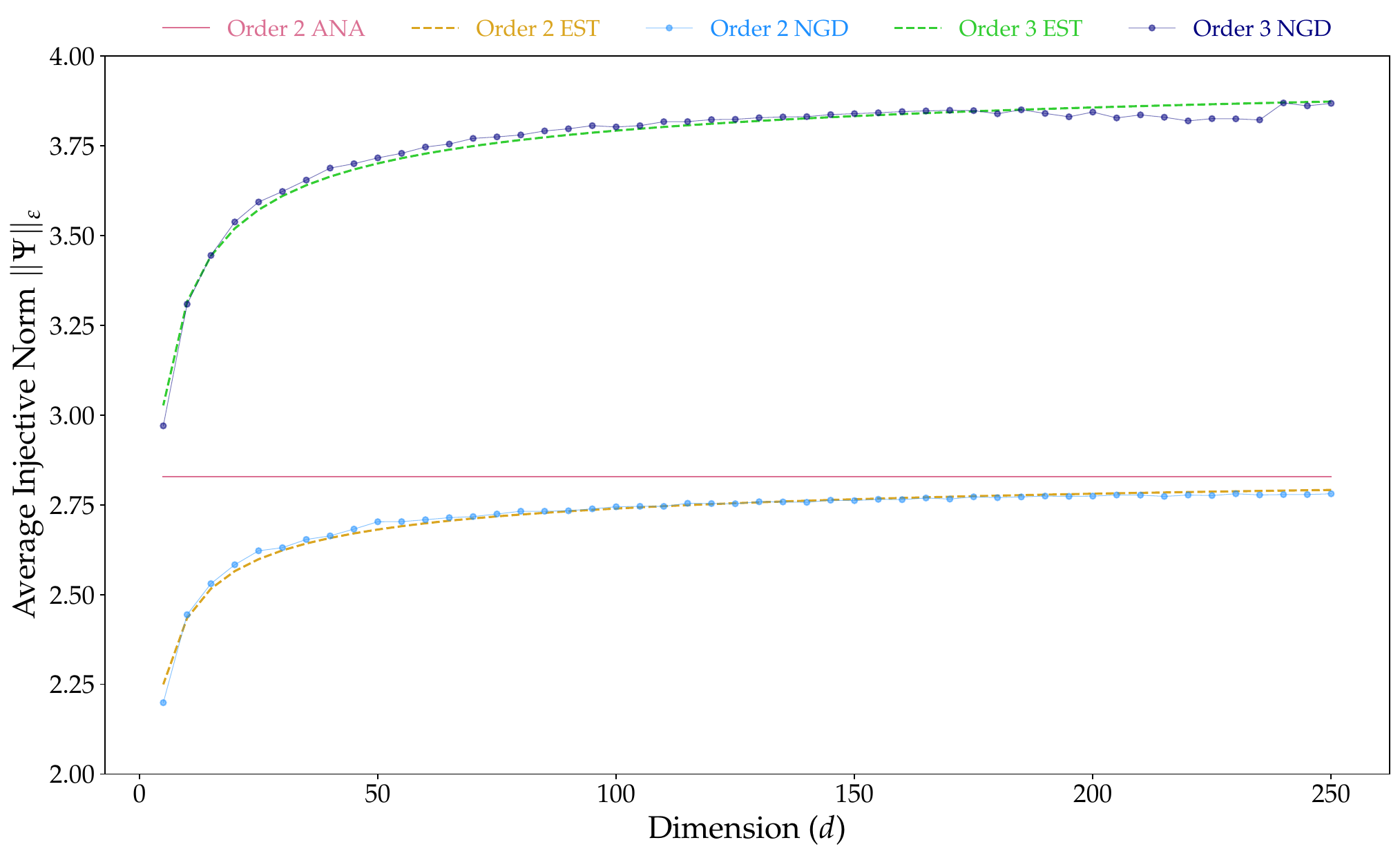}
     \caption{Non-symmetrized complex Gaussian tensors.}
     \label{fig:c_non_gauss_inj}
    \end{subfigure}
    \end{figure}
    
    \begin{figure}[H]\ContinuedFloat
    \centering
    \begin{subfigure}[b]{0.9\textwidth}
    \centering
    \includegraphics[width=\textwidth]{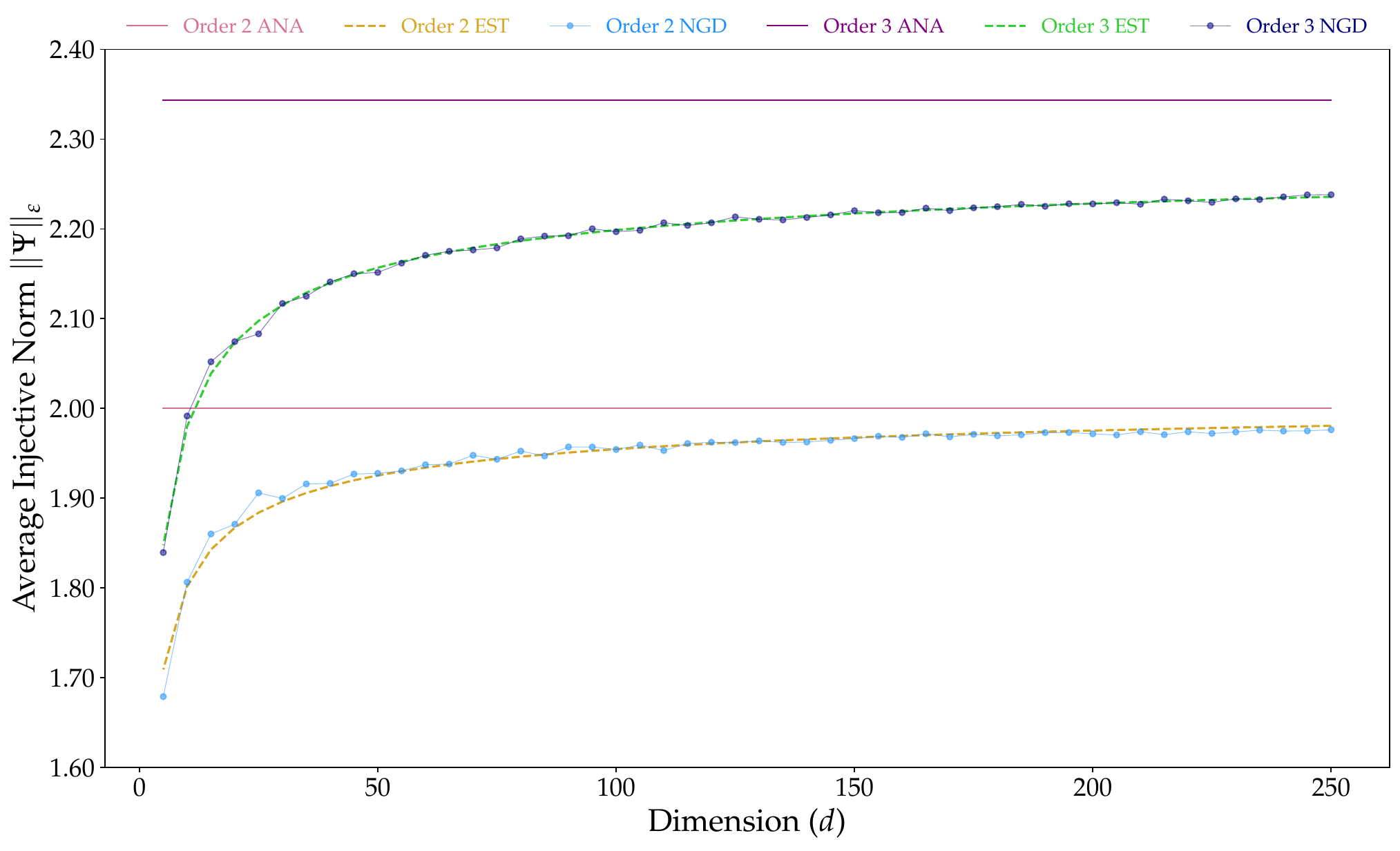}
    \caption{Symmetrized complex Gaussian tensors.}
    \label{fig:c_sym_gauss_inj}
    \end{subfigure}
    \caption{We approximate the average value of the injective norm of both symmetrized and non-symmetrized complex Gaussian tensors using our NGD algorithm. We estimate the asymptotes $C_0$ by fitting geometric series on the first differences of these data points. We then fit function estimators of the form \(C_0 + C_1 d^\alpha\), where $C_0$ is fixed as the known or conjectured asymptotic value. The values of \(C_1\) and \(\alpha\) are estimated by minimizing the squared error and are given in Table \ref{table:4}.}
    \label{fig:c_gauss_inj}
\end{figure}

\newpage
\subsection{The effect of symmetrization on multipartite entanglement}

In this section, we want to understand how symmetrizing a multipartite pure state affects the amount of multipartite entanglement present in it. More precisely, given a multipartite pure state \(\hat\Psi\in(\mathbb C^d)^{\otimes n}\), we define its symmetrization
\[ \Psi_S:=\frac{1}{n!}\sum_{\sigma\in S_n} \sigma.\hat\Psi, \]
where \(\sigma\in S_n\) acts on \(\hat\Psi\) by permuting the tensor factors, and the corresponding symmetric multipartite pure state \(\hat\Psi_S=\frac{\Psi_S}{\|\Psi_S\|}\). \(\Psi_S\) is the projection of \(\hat\Psi\) onto the symmetric subspace of \((\mathbb C^d)^{\otimes n}\), it is in general subnormalized. We would then like to compare the multipartite entanglement of \(\hat\Psi\) and \(\hat\Psi_S\). Intuitively, one would think that \(\hat\Psi_S\) is generically less entangled than \(\hat\Psi\), i.e.,
\begin{equation} \label{eq:ent-sym}
\| \hat{\Psi}_S \|_\varepsilon \geq \|\hat{\Psi}\|_\varepsilon.
\end{equation}
Note that Equation \eqref{eq:ent-sym} does not hold true for all multipartite pure states. For instance, \(\hat\Psi=|12\rangle\in(\mathbb C^2)^{\otimes 2}\) is separable, while its normalized symmetrization \(\hat\Psi_S=\frac{1}{\sqrt{2}}(|12\rangle+|21\rangle)\) is maximally entangled. However, we expect that, for large \(d\), multipartite pure states in \((\mathbb C^d)^{\otimes n}\) typically satisfy Equation \eqref{eq:ent-sym} when the injective norm is averaged over several samples of random states $\hat\Psi \in \mathbb({C^d})^{\otimes n}$.\\

We employ our NGD algorithm to quantitatively study the effect of symmetrization on the amount of entanglement present in random multipartite pure quantum states with a Gaussian distribution. We do so by computing the average normalized injective norm of both symmetrized and non-symmetrized complex Gaussian tensors. Indeed, we recall that \(\|\hat{\Psi}\|_\varepsilon = 2^{-\GME(\Psi)/2}\) is equal to \(1\) only for product states \(\hat{\Psi}\) and approaches \(0\) as \(d\) grows for highly entangled states.\\

Figures \ref{fig:c_non_gauss_ratio} and \ref{fig:c_sym_gauss_ratio} show that entanglement increases with the dimension \(d\) of the tensor, regardless of symmetry. Also, the value of \(\|\hat{\Psi}\|_\varepsilon\) and \(\GME(\Psi)\) for order \(2\) tensors remains the same, regardless of symmetry, owing to the choice of normalization factor described in Section \ref{sec:r_sym_gauss}. However, for order \(3\) tensors, symmetrization decreases the amount of entanglement in the tensor and increases the value of \(\|\hat{\Psi}\|_\varepsilon\). The plots in Figure \ref{fig:c_mix_gauss_ratio} illustrate that for order \(3\) tensors, symmetrization results in a larger value of \(\|\hat{\Psi}\|_\varepsilon\), i.e.,~symmetrized tensors are less entangled than their non-symmetrized counterparts. Hence, we can conclude that projecting random Gaussian tensors onto the symmetric subspace of \(\left(\mathbb{C}^d\right)^{\otimes n}\) results in less entangled tensors, on average.

\begin{figure}[H]
\centering
    \begin{subfigure}[b]{0.9\textwidth}
         \centering
         \includegraphics[width=\textwidth]{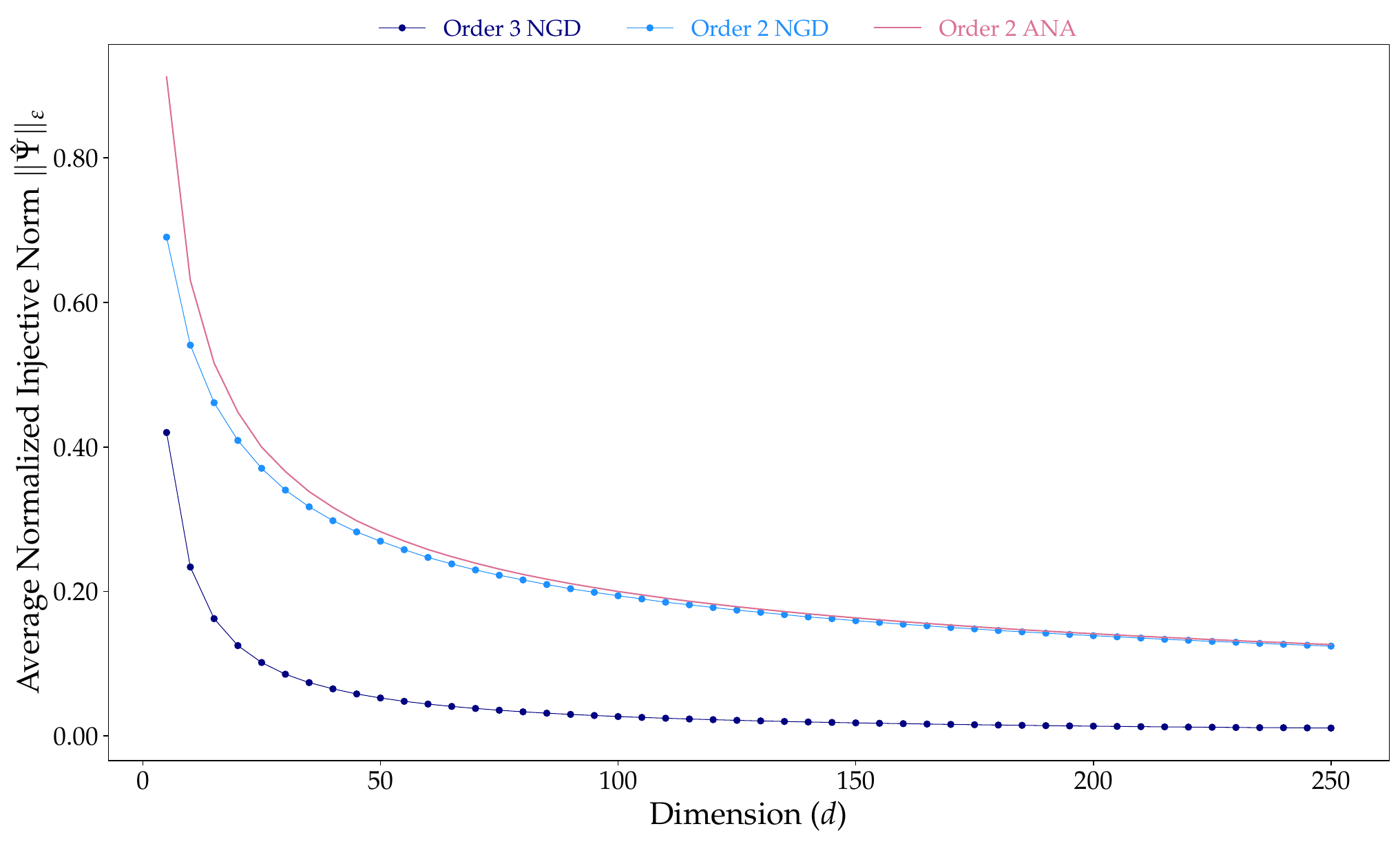}
         \caption{Non-symmetrized complex Gaussian tensors.}
         \label{fig:c_non_gauss_ratio}
     \end{subfigure}
\end{figure}
\begin{figure}[H]\ContinuedFloat
\centering
     \begin{subfigure}[b]{0.9\textwidth}
         \centering
         \includegraphics[width=\textwidth]{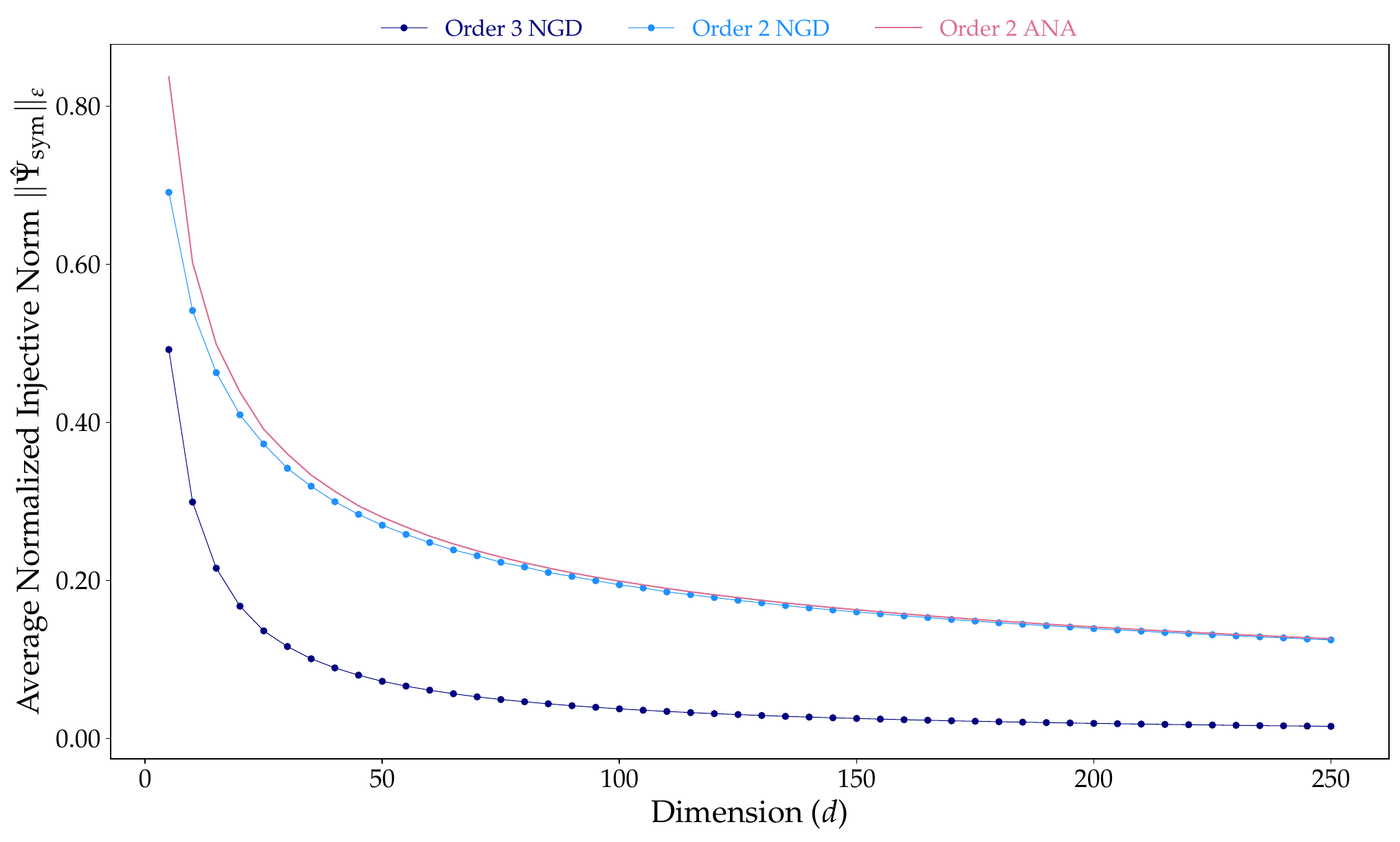}
         \caption{Symmetrized complex Gaussian tensors.}
         \label{fig:c_sym_gauss_ratio}
     \end{subfigure}
\end{figure}
\begin{figure}[H]\ContinuedFloat
\centering
     \begin{subfigure}[b]{0.9\textwidth}
         \centering
         \includegraphics[width=\textwidth]{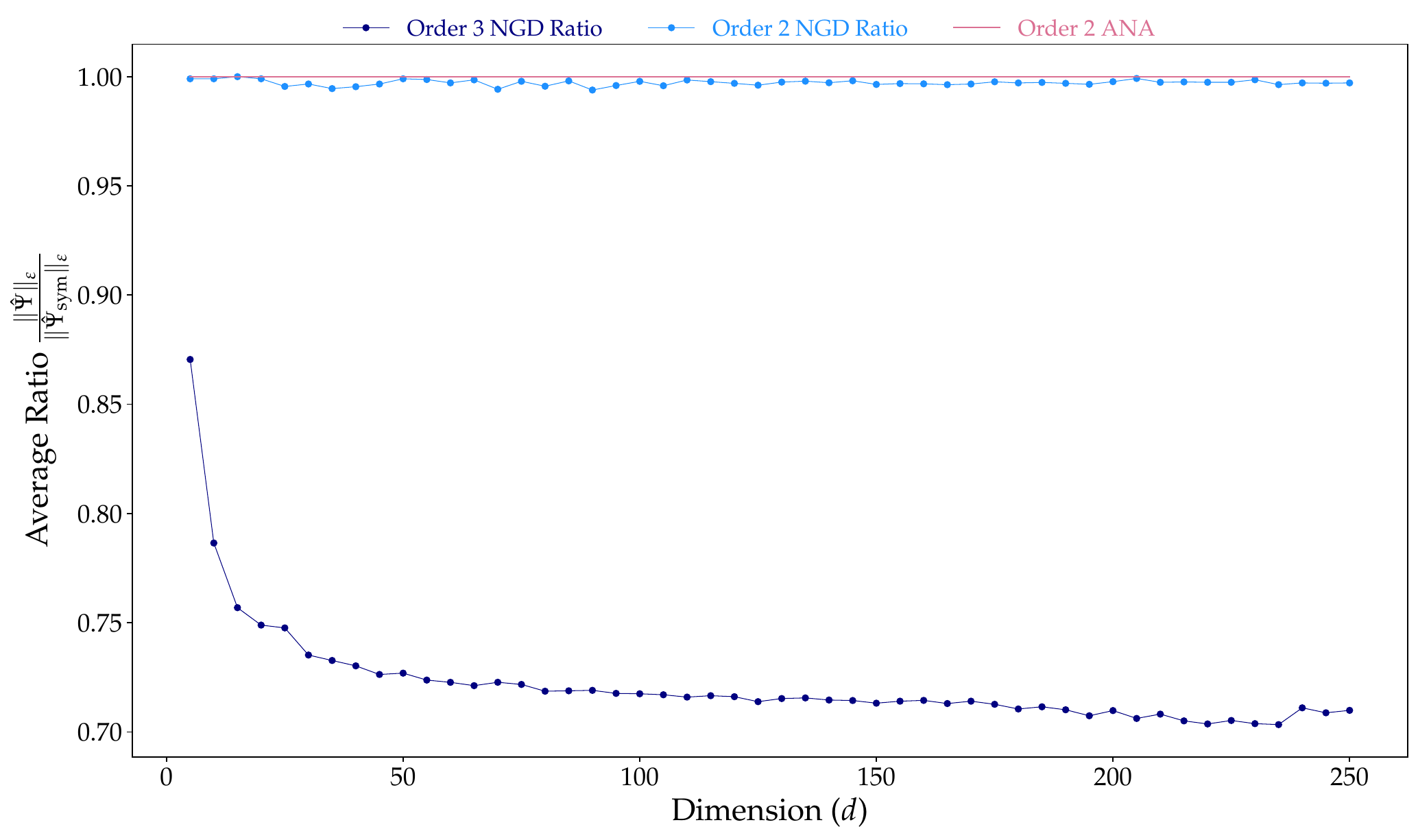}
         \caption{The ratio of the normalized injective norm of non-symmetrized complex Gaussian tensors to that of their symmetrized counterparts.}
         \label{fig:c_mix_gauss_ratio}
     \end{subfigure}
    \caption{Entanglement in both symmetrized and non-symmetrized complex Gaussian tensors increases with increasing \(d\). The value of \(\|\hat{\Psi}\|_\varepsilon\) for symmetrized Gaussian tensors is greater than (for order \(3\) and above) or equal to (for order \(2\)) that for non-symmetrized Gaussian tensors of the same order and dimension. This indicates that symmetrizing random Gaussian tensors reduces entanglement, on average.}
    \label{fig:c_gauss_ratio}
\end{figure}


\section{Random matrix product states}
\label{sec:MPS}
In this section, we look at a model of random multipartite pure quantum states, which might be more interesting than uniformly distributed states from a physical point of view, namely, random matrix product states (MPS). MPS are used extensively in quantum condensed matter physics as an ansatz for the ground states of gapped local Hamiltonians on 1D systems \cite{hastings2006solving,landau2015polynomial}. Studying typical properties of MPS is a research direction that has recently attracted a lot of attention, as it ties closely with studying typical properties of physically relevant many-body quantum systems. The quantities that have been extensively studied up to now include the generic amount of entanglement between a bipartition of the sites \cite{garnerone2010typicality,collins2013matrix,gonzalezguillen2018spectral,haferkamp2021emergent} and the generic amount of correlations between two distant sites \cite{lancien2022correlation}. Similar to the case of complex Gaussian random tensors, we go beyond the bipartite paradigm and present the first estimates on the genuinely multipartite entanglement of random MPS.\looseness-1

\subsection{Gaussian matrix product states}
To construct a random MPS \(X \in \mathbb{C}^{d_1} \otimes \mathbb{C}^{d_2} \otimes \cdots \otimes \mathbb{C}^{d_n} \) with periodic boundary conditions, we start by sampling \(n\) local tensors \(A^{(k)} \in \mathbb{C}^{q_k} \otimes \mathbb{C}^{d} \otimes \mathbb{C}^{q_{k+1}}\), for \(k \in \{1, 2, \ldots, n\}\), with \(q_{n+1} = q_1\). The local dimensions \(d_k\) and \(q_k\) are called the physical and bond dimensions, respectively.

\begin{equation}\label{eqn:mps_sample}
A^{(k)}_{i_1, s, i_2} \sim \mathcal{N}\left( 0, \frac{2}{d_k\sqrt{q_kq_{k+1}}} \right) \;\forall\; i_1 \in \{1, 2, \ldots , q_k\}, s \in \{1, 2, \ldots , d_k\}, i_2 \in \{1, 2, \ldots , q_{k+1}\}.
\end{equation}

We sample the real and imaginary parts of each entry of the local tensors as described in Equation \eqref{eqn:mps_sample}, and normalize by a factor of \(\sqrt{2}\). We then contract the local tensors over the internal bond dimensions and identify the last index with the first index.

\begin{equation}\label{eqn:mps}
    X_{s_1, s_2, \cdots , s_n} = \sum_{i_1,i_2,\cdots,i_n} A^{(1)}_{i_1 s_1 i_2} A^{(2)}_{i_2 s_2 i_3} \cdots A^{(n)}_{i_n s_n i_1}.
\end{equation}

When the local tensors \(A^{(k)}\) are sampled as described in Equation \eqref{eqn:mps_sample}, independently from one another, the expected value of the squared Euclidean norm of the corresponding Gaussian MPS is simply \(d_1\times\cdots\times d_n\times q_1\times\cdots\times q_n\times\frac{2}{d_1\sqrt{q_1q_2}}\times\cdots\times\frac{2}{d_n\sqrt{q_nq_1}}=2^n\). Translation-invariance can be incorporated in this model of random MPS by repeating the same local tensor \(A^{(k)}\) on all sites, with all \(q_k = q\), \(d_k = d\). In this case, the expected value of the squared Euclidean norm of the translation-invariant Gaussian MPS is \(2^n+o(1)\) as \(d,q\rightarrow\infty\).\looseness-1\\

In what follows, we focus on presenting estimates on the normalized injective norm in the following cases: random MPS with periodic boundary conditions, constructed from either independent or repeated complex Gaussian tensors on each site. We will present the results obtained for such random tensors of order \(3\). As already explained, computing the injective norm of order \(2\) tensors is not particularly challenging, while all the theoretical and computational challenges (NP-hardness) encountered for higher order tensors are already encompassed in the order \(3\) case.\\

An important observation is that the normalized injective norm of a Gaussian MPS, unlike the one of a Gaussian tensor, does not necessarily go to \(0\) as \(d\to \infty\). It indeed depends on both parameters \(d\) and \(q\), in a way that is quite complicated to grasp. The goal of the subsequent analysis is to try and understand better this dependence on \(d\) and \(q\), in different asymptotic regimes. \\

In the work in progress \cite{lancien2022injective}, it is proven that the injective norm of a Gaussian MPS in \((\mathbb C^d)^{\otimes n}\) with bond dimension \(q\) (either translation-invariant or not) is of order \(\max\left(\frac{1}{\sqrt{d^{n-1}}},\frac{1}{\sqrt{q^{n-1}}}\right)\), as \(d,q\rightarrow\infty\) (up to a potential \(n\)-dependent multiplicative factor). It is not clear how this joint asymptotic scaling in \(d\) and \(q\) could be numerically confirmed in a reliable way. It would indeed require to fit the asymptotic injective norm as \(d\) and \(q=q(d)\) grow at various relative speeds. Instead, we focus on the two extreme cases where either \(q\) is fixed and \(d\) grows or \(d\) is fixed and \(q\) grows.\looseness-1


\subsection{Scaling with physical dimension for fixed bond dimension}
Similar to the Gaussian tensors, we estimate the numerical lower bounds on the asymptotic injective norm of order \(3\) Gaussian MPS with a fixed bond dimension \(q\), as \(d \rightarrow \infty\). The results are presented in Figures \ref{fig:mps_bf_non_inv} (for independent local tensors) and \ref{fig:mps_bf_inv} (for repeated local tensors). We use the same approach of fitting a geometric series of the form \((C_0r^k)_{k\in\mathbb N}\) on the first differences of the data points. We report the values of the asymptotes in Table \ref{table:mps_asymp_bf}. 

\vspace{-2mm}
\begin{figure}[H]
    \centering
    \begin{subfigure}{0.9\textwidth}
    \centering
    \includegraphics[width=\textwidth]{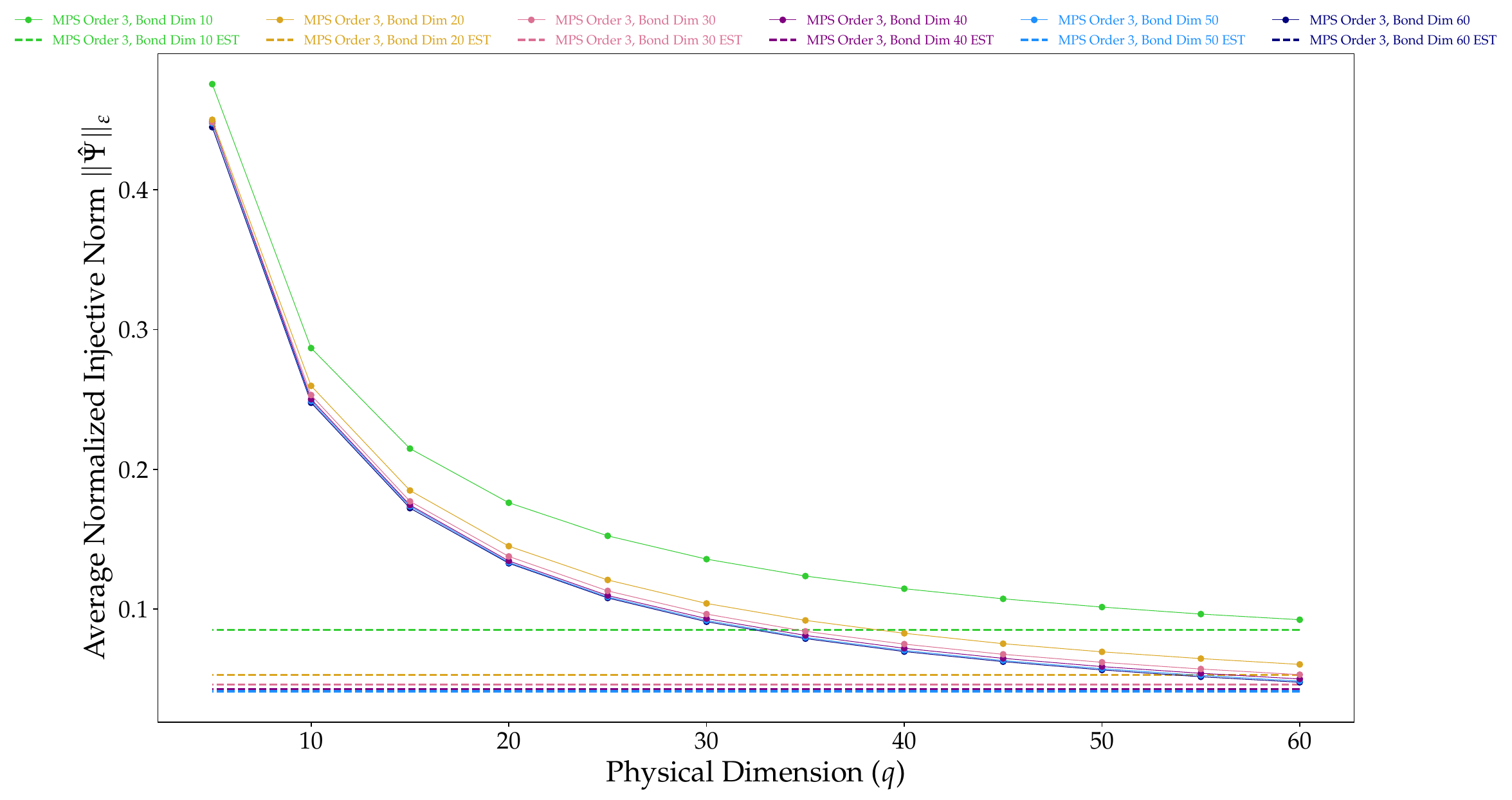}\vspace{-1mm}
    \caption{Gaussian MPS without translation-invariance.}
    \label{fig:mps_bf_non_inv}
    \end{subfigure}
    \end{figure}\vspace{-4mm}
    \begin{figure}[H]\ContinuedFloat
    \centering
    \begin{subfigure}{0.9\textwidth}
    \centering
    \includegraphics[width=\textwidth]{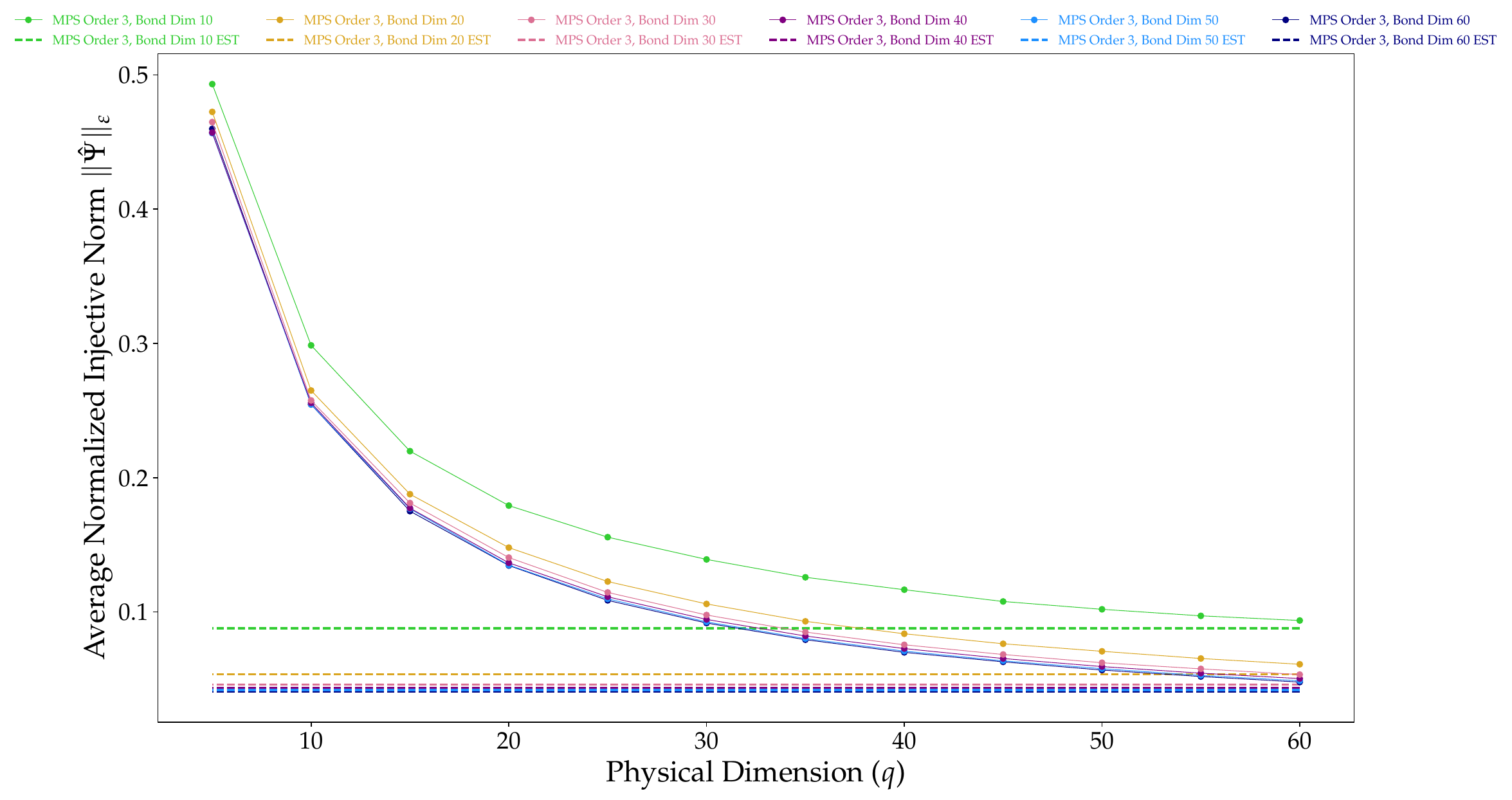}\vspace{-1mm}
    \caption{Gaussian MPS with translation-invariance.}
    \label{fig:mps_bf_inv}
    \end{subfigure}
    \caption{For both cases (with and without translation-invariance), for a fixed \(q\), as \(d\) increases, the value of the normalized injective norm decreases, indicating that entanglement increases with \(d\). Further, we fit a geometric series of the form \((C_0r^k)_{k\in\mathbb N}\) on the first differences of the data points obtained using our NGD algorithm. The estimated values of the asymptotes are given in Table \ref{table:mps_asymp_bf}.}
\end{figure}

\vspace{-6mm}
\begin{table}[t]
\centering
\footnotesize{
\begin{tabular}{lcr}
\hline
\multicolumn{1}{c}{Translation-invariance} &
  Bond Dimension & 
  Estimated asymptote\\\hline
\multirow{6}{*}{With}  & \multirow{1}{*}{10}    & \(0.087853\) \\
                    & \multirow{1}{*}{20}    & \(0.053398\) \\
                    & \multirow{1}{*}{30}    & \(0.045909\) \\
                    & \multirow{1}{*}{40}    & \(0.043447\) \\
                    & \multirow{1}{*}{50}    & \(0.041846\) \\
                    & \multirow{1}{*}{60}    & \(0.040545\) \\  \hline
\multirow{6}{*}{Without}  & \multirow{1}{*}{10}    & \(0.085244\) \\
                    & \multirow{1}{*}{20}    & \(0.053005\) \\
                    & \multirow{1}{*}{30}    & \(0.046096\) \\
                    & \multirow{1}{*}{40}    & \(0.042909\) \\
                    & \multirow{1}{*}{50}    & \(0.041235\) \\
                    & \multirow{1}{*}{60}    & \(0.040860\) \\  \hline
\end{tabular}
}
\caption{Estimates for the asymptotes with fixed bond dimensions, as $d\rightarrow \infty$, obtained by fitting a geometric series of the form \((C_0r^k)_{k\in\mathbb N}\) on the first differences of the data points obtained using our NGD algorithm for Gaussian MPS with periodic boundary conditions, with and without translation-invariance.\looseness-1}
\label{table:mps_asymp_bf}
\end{table}

\subsection{Scaling with bond dimension for fixed physical dimension}

Similar to the case with fixed bond dimensions, we also estimate the asymptotes for fixed physical dimensions $d$, as $q\rightarrow\infty$, by fitting a geometric series of the form \((C_0r^k)_{k\in\mathbb N}\) on the first differences of the data points obtained using our NGD algorithm for Gaussian MPS with periodic boundary conditions, with and without translation-invariance.  We present these estimates in Table \ref{table:mps_asymp_df}.\\

We are interested in comparing the amount of entanglement in a Gaussian MPS and its corresponding Gaussian tensor. Before explaining the correspondence between Gaussian MPS and Gaussian tensors, we state our conjecture.
\newpage
\begin{conj}
If the physical dimension \(d\) is fixed, then as the bond dimension \(q\) increases, the value of the normalized injective norm of a Gaussian MPS in \(\left(\mathbb{C}^{d}\right)^{\otimes n}\) converges to that of a corresponding Gaussian tensor in \(\left(\mathbb{C}^{d}\right)^{\otimes n}\).\\    
\end{conj}

The latter is either a non-symmetrized Gaussian tensor, in the case where the MPS has independent local tensors, or a cyclically symmetrized Gaussian tensor (see precise definition below),  in the case where the MPS has repeated local tensors. In what follows, we explain our choice of correspondence between MPS and Gaussian tensors, along with numerical evidence to support our claim.

\subsubsection{Comparing non-translation-invariant MPS to non-symmetrized Gaussian tensors}
We compare MPS in \(\left(\mathbb{C}^{d}\right)^{\otimes n}\) with \(n\) sites, each having a physical dimension \(d\) and bond dimension \(q\), with periodic boundary conditions and independent Gaussian local tensors (i.e.,~without translation-invariance) to non-symmetrized Gaussian tensors in \(\left(\mathbb{C}^{d}\right)^{\otimes n}\). Since all entries for each local tensor in such an MPS are drawn independently and identically from the same distribution, and all local tensors contribute equally to the construction (periodic boundary conditions), we consider this case to be the closest to a non-symmetrized Gaussian tensor.

\subsubsection{Comparing translation-invariant MPS to cyclically symmetrized Gaussian tensors}
Further, we compare MPS in \(\left(\mathbb{C}^{d}\right)^{\otimes n}\) with \(n\) sites, each having a physical dimension \(d\) and bond dimension \(q\), with periodic boundary conditions and repeated Gaussian local tensors (i.e.,~with translation-invariance) to cyclically symmetrized Gaussian tensors in \(\left(\mathbb{C}^{d}\right)^{\otimes n}\). Since all sites in such an MPS are identical, the entries of the final MPS are invariant under cyclic permutations of the indices. This makes the cyclically symmetrized Gaussian tensors (whose precise definition is given below) a natural choice for comparison with such MPS.\\

Similar to the symmetrized Gaussian tensors described in Section \ref{sec:r_sym_gauss}, a cyclically symmetrized complex Gaussian tensor \(X_c \in \left(\mathbb{C}^d\right)^{\otimes n}\) can be formed by projecting a complex Gaussian tensor \(X\in \left(\mathbb{C}^d\right)^{\otimes n}\) onto the cyclically symmetric subspace of \(\ \left(\mathbb{C}^d\right)^{\otimes n}\). So instead of averaging over all permutations of the auxiliary tensor's axes, we average only over cyclic permutations of the axes.
\begin{equation} \label{}
    X_{c_{i_1, i_2,\ldots , i_n}} = \frac{1}{n}\sum_{\sigma\in C_n}X_{i_{\sigma(1)}, i_{\sigma(2)},\ldots , i_{\sigma(n)}} \;\forall\; i_{\sigma(j)} \in \{1, 2, \ldots , d\} \;\forall\; j \in \{1, 2, \ldots , n\}.
\end{equation}
where $C_n$ is the cyclic permutation group of \(n\) elements. The cyclically symmetrized tensor \(X_c\) has the real and imaginary parts of its entries sampled from \(\mathcal{N}\big(0, \frac{2}{dn}\big)\) for locations with no repeated indices. Its expected Euclidean norm is asymptotically (as \(d\rightarrow\infty\)) scaled by a factor \(\frac{1}{\sqrt{n}}\) as compared to the Euclidean norm of the non-symmetrized tensor.\\

As we show in Figures \ref{fig:mps_vs_non_gauss} and \ref{fig:mps_vs_cyc_gauss}, when the local dimension \(d\) is fixed, the value of the normalized injective norm of a Gaussian MPS saturates, as \(q \to \infty\), towards that of a corresponding Gaussian tensor with the same local dimension \(d\). One can verify this by comparing the asymptotes, for a given \(d\), with the value of the estimated injective norm of Gaussian tensors with the same dimension.\looseness-1

\begin{table}[H]
\centering
\footnotesize{
\begin{tabular}{lcrr}
\hline
\multicolumn{1}{c}{Translation-invariance} &
  Bond Dimension & 
  Estimated asymptote &
  Estimated injective norm of Gaussian tensors\\\hline
\multirow{6}{*}{With}  & \multirow{1}{*}{10}    & \(0.256331\) & \(0.240307\)\\
                    & \multirow{1}{*}{20}    & \(0.134856\) & \(0.126094\)\\
                    & \multirow{1}{*}{30}    & \(0.091259\) & \(0.085945\)\\
                    & \multirow{1}{*}{40}    & \(0.069404\) & \(0.065318\)\\
                    & \multirow{1}{*}{50}    & \(0.056270\) & \(0.052702\)\\
                    & \multirow{1}{*}{60}    & \(0.047135\) & \(0.044232\)\\  \hline
\multirow{6}{*}{Without}  & \multirow{1}{*}{10}    & \(0.245287\) & \(0.233780\)\\
                    & \multirow{1}{*}{20}    & \(0.132823\) & \(0.125040\)\\
                    & \multirow{1}{*}{30}    & \(0.090227\) & \(0.085408\)\\
                    & \multirow{1}{*}{40}    & \(0.069452\) & \(0.065196\)\\
                    & \multirow{1}{*}{50}    & \(0.055968\) & \(0.052555\)\\
                    & \multirow{1}{*}{60}    & \(0.047188\) & \(0.044149\)\\  \hline
\end{tabular}
}
\caption{Estimates for the asymptotes for some fixed physical dimension $d$, as $q\rightarrow\infty$, obtained by fitting a geometric series of the form \((C_0r^k)_{k\in\mathbb N}\) on the first differences of the data points obtained using our NGD algorithm for Gaussian MPS with periodic boundary conditions, with and without translation-invariance. Further, we compare these asymptotes with the estimated injective norms of non-symmetrized and cyclically symmetrized Gaussian tensors of the same dimension. This provides numerical evidence to support our claim that as the bond dimension increases, the injective norm of a Gaussian MPS converges to that of a Gaussian tensor with similar symmetries and physical (local) dimension.}
\label{table:mps_asymp_df}
\end{table}

\vspace{-4mm}
\begin{figure}[H]
    \centering
    \begin{subfigure}[b]{\textwidth}
    \centering
    \includegraphics[width=\textwidth]{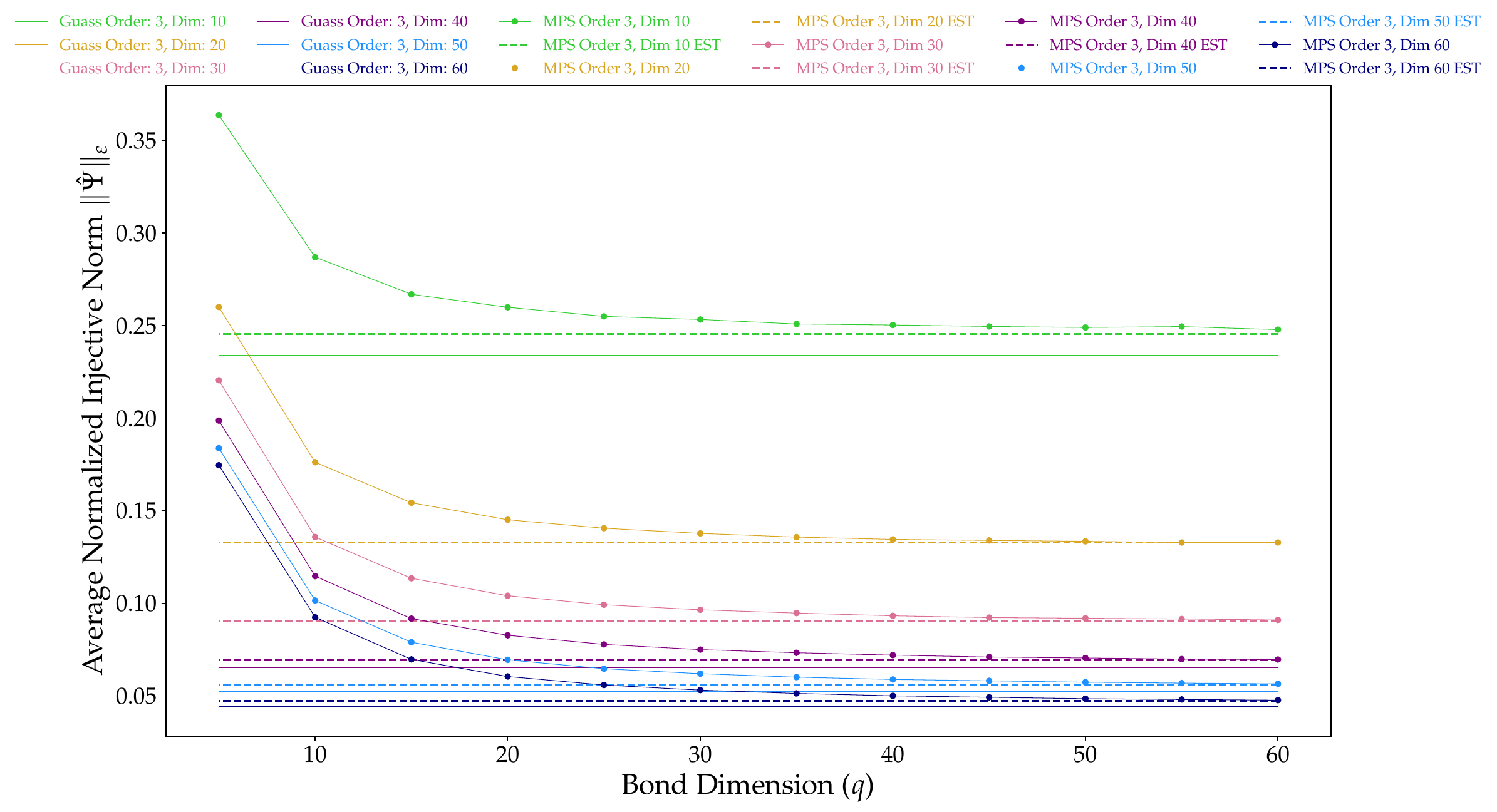}
    \caption{MPS without translation-invariance compared to non-symmetrized complex Gaussian tensors.}
    \label{fig:mps_vs_non_gauss}
    \end{subfigure}
\end{figure}
\begin{figure}[H]\ContinuedFloat
    \centering
    \begin{subfigure}[b]{\textwidth}
    \centering
    \includegraphics[width=\textwidth]{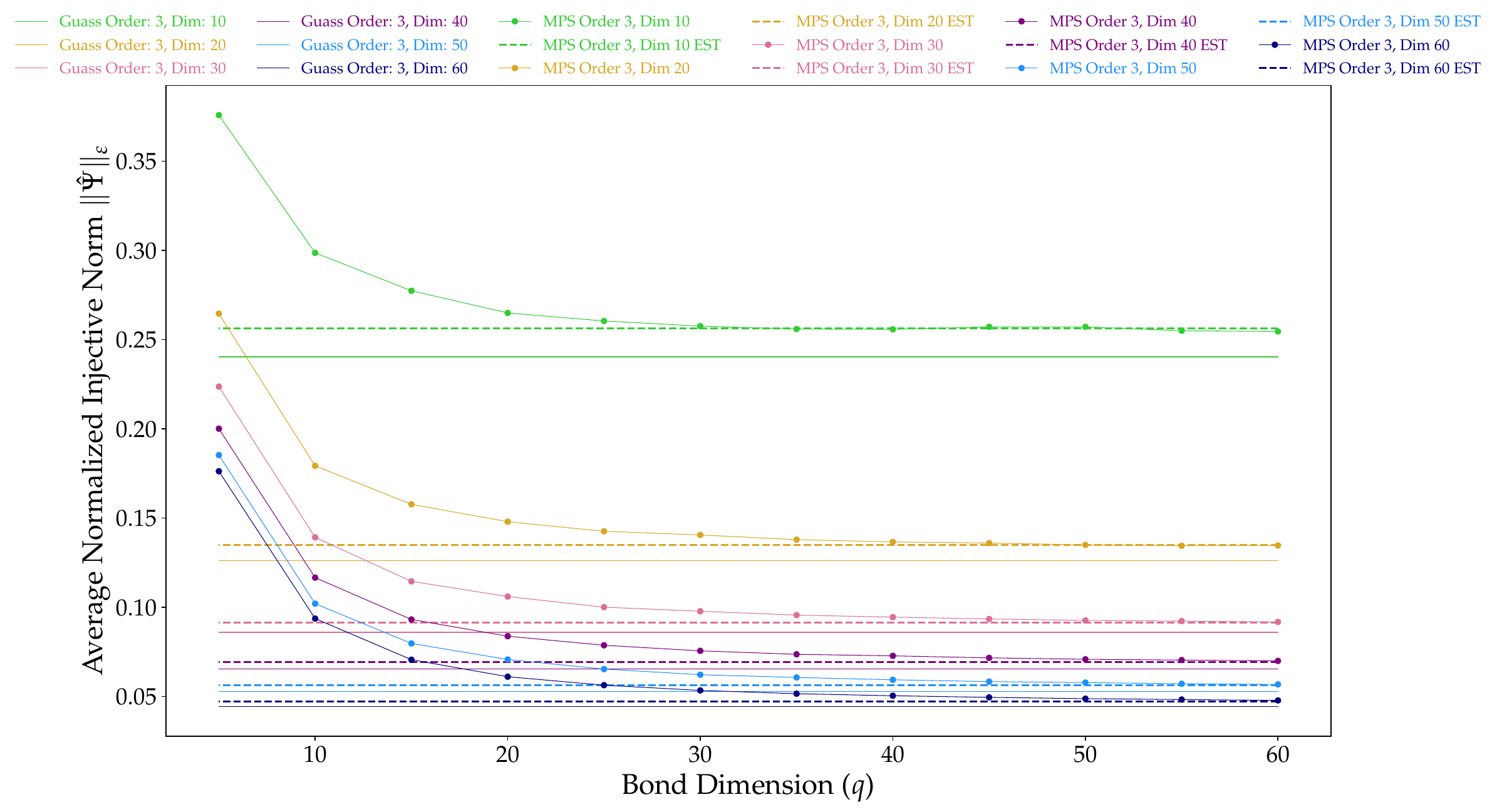}
    \caption{MPS with translation-invariance compared to cyclically symmetrized complex Gaussian tensors.}
    \label{fig:mps_vs_cyc_gauss}
    \end{subfigure}
    \caption{Similar to the case with fixed bond dimension \(q\) and increasing physical dimension \(d\), when we fix the physical dimension \(d\) and increase the bond dimension \(q\), the entanglement increases. We compare order \(3\) Gaussian MPS with and without translation-invariance, with cyclically symmetrized and non-symmetrized complex Gaussian tensors with the same physical (local) dimension \(d\), respectively. In both cases, for a given \(d\), as \(q \to \infty\), the amount of entanglement present in the Gaussian MPS approaches that of the corresponding complex Gaussian tensor, hence supporting our conjecture.}
\end{figure}

\section{Summary and perspectives}

In this work, we develop an algorithm for estimating the injective norm of tensors, which is equivalent to estimating the geometric measure of entanglement of the corresponding multipartite quantum state. This is a notoriously difficult problem (NP-hard for tensors of order \(3\) or beyond), and several heuristic algorithms have been considered in the literature. The novel algorithm we propose is based on a gradient descent approach. We first study its performance on normalized tensors with a large number of parties, whose injective norm is known, for which we observe that it performs at par with the usual methods based on alternating least squares (ALS). However, when dealing with tensors having large local dimension, it is usually necessary to work with non-normalized tensors because the injective norm of such tensors is typically much smaller than their Euclidean norm. Our algorithm substantially outperforms the ALS-based methods on such non-normalized inputs, with a disparity in performance that scales polynomially with the local dimension.\\

We then use our algorithm to estimate the injective norm of various random tensors of order \(2\) and \(3\): Gaussian tensors (real or complex and symmetric or not), Gaussian MPS (translation-invariant or not and cyclically symmetric or not). 
In particular, our results show that all corresponding random tripartite pure states are typically genuinely multipartite entangled, and indeed, quite strongly so. If they exhibited only bipartite entanglement (even approximately), their injective norm would decay slowly than \(\frac{1}{\sqrt{d}}\). However, we observe that it instead scales as \(\frac{1}{d}\), indicating a nearly maximal amount of tripartite entanglement. To the best of our knowledge, this is the first such observation for MPS. Although extensive studies have examined bipartite entanglement and correlations in MPS, essentially nothing was previously known about the amount of genuine multipartite entanglement in these physically relevant, structured tensors.\\

Finally, we mention that after this work was completed, two papers whose goal was to analytically estimate the asymptotic injective norm of non-symmetrized Gaussian tensors came out \cite{dartois2024,sasakura2024signed}. The work \cite{sasakura2024signed} shows that, for order \(3\) tensors, the numerical value that we obtain for the asymptotic injective norm of complex non-symmetrized Gaussian tensors, namely \(4.14\), is very close to the analytical value \(4\). Similarly, the work \cite{dartois2024} gives an analytical upper bound of \(4\) for the asymptotic injective norm of real non-symmetrized, which is compatible with (and very close to) the numerical value \(3.95\) that we obtain in this case. Finally, \cite{dartois2024} provides strong evidence supporting our conjecture that real and complex Gaussian tensors have the same asymptotic injective norm (in the non-symmetrized case). Indeed, the authors find the same analytical upper bounds in real and complex cases and they conjecture that these upper bounds are both tight. In addition to corroborating the performance of our algorithms, these two works confirm that the question of estimating the injective norm of random tensors is interesting in a variety of fields, such as spin glasses for \cite{dartois2024} and quantum gravity for \cite{sasakura2024signed}.\\

To conclude, our work achieves two equally important objectives. First, it introduces an efficient algorithm for estimating the injective norm of tensors, which we expect to be valuable to various communities that utilize tensor norms--for instance, in quantum information theory for estimating the geometric measure of entanglement of multipartite pure states, in statistical physics for estimating the ground state energy of spin models, among other applications. Second, it provides novel numerical insights into the typical amount of genuine multipartite entanglement present in several natural classes of pure states, for which analytical results have so far remained elusive. We hope our results motivate future efforts towards deriving analytical bounds, if not exact values, for these injective norms.

\bigskip

\noindent\textbf{Acknowledgements.} K.F.~thanks Rishika~Bhagwatkar, Akshay Kulkarni, Aditya Shirwatkar, Prakrut Kotecha, and Aaditya Rudra for their timely assistance with running and monitoring the experiments. The authors were supported by the ANR project \href{https://esquisses.math.cnrs.fr/}{ESQuisses}, grant number ANR-20-CE47-0014-01. K.F.~acknowledges support from a \href{https://nanox-toulouse.fr/}{NanoX} project grant. C.L.~and I.N.~acknowledge support by the ANR project \href{https://www.math.univ-toulouse.fr/~gcebron/STARS.php}{STARS}, grant number ANR-20-CE40-0008, and by the PHC programs \emph{Sakura} (Random Matrices and Tensors for Quantum Information and Machine Learning) and \emph{Procope} (Entanglement Preservation in Quantum Information Theory). C.L.~was also supported by the ANR project \href{https://qtraj.math.cnrs.fr/}{QTraj}, grant number ANR-20-CE40-0024-01, while I.N.~was also supported by the PHC Program \emph{Star} (Applications of random matrix theory and abstract harmonic analysis to quantum information theory). 



\bibliographystyle{quantum} 
\bibliography{refs}
\newpage
\appendix
\section{Estimating the injective norm of deterministic multipartite pure states}\label{appendix:det_states}
As a part of our initial benchmarking experiments, we test the ALS and NGD algorithms on deterministic states for which the value of GME is well-known \cite{zhu2010additivity}. Here, we present mainly two types of states, namely generalized Dicke states and antisymmetric basis states.

\subsection{Generalized Dicke states}
Dicke states have found several applications in quantum communication and quantum networking \cite{kiesel2007photon, prevedal2009qubit} along with some typical Dicke states being realized using trapped atomic ion systems \cite{hume2009dicke}. Given an orthonormal basis \(\{e_1,e_2,\ldots,e_d\}\) of \(\mathbb C^d\), the associated generalized Dicke state \(\Psi_D(n,d, \vec{k}) \in \left(\mathbb{C}^d\right)^{\otimes n}\) is defined as

\begin{equation}\label{eqn:dicke}
    \begin{aligned}
&\Psi_D(n,d, \vec{k}):=\frac{1}{\sqrt{C_{n,d, \vec{k}}}} \; \sum_{\sigma \in S_{n,d, \vec{k}}} \overbrace{e_{\sigma(1)}\otimes\cdots\otimes e_{\sigma(1)}}^{k_1} \otimes \overbrace{e_{\sigma(2)}\otimes \cdots \otimes e_{\sigma(2)}}^{k_2} \otimes \cdots \otimes \overbrace{e_{\sigma(d)} \otimes \cdots \otimes e_{\sigma(d)}}^{k_d} \\
&\vec{k}:=\left(k_1, k_2, \ldots, k_d\right), \quad \sum_{j=1}^d k_j=n,
\end{aligned}
\end{equation}

where \(S_{n,d, \vec{k}}\) denotes the set of all distinct permutations of the set \(\{\overbrace{1,\ldots,1}^{k_1},\overbrace{2,\ldots,2}^{k_2},\ldots,\overbrace{d\ldots,d}^{k_d}\}\) and \(C_{n,d, \vec{k}} = \frac{n!}{\prod_{j=1}^d k_j!}\) is the normalization factor. The analytical value for the GME of such states is known \cite{zhu2010additivity} and is given by

\begin{equation}
\GME\left(\Psi_D(n,d, \vec{k})\right) = \log \left[\frac{1}{C_{n,d, \vec{k}}} \prod_{j=1}^d\left(\frac{n}{k_j}\right)^{k_j}\right] .
\end{equation}

\begin{figure}[H]
    \centering
    \includegraphics[width=0.9\textwidth]{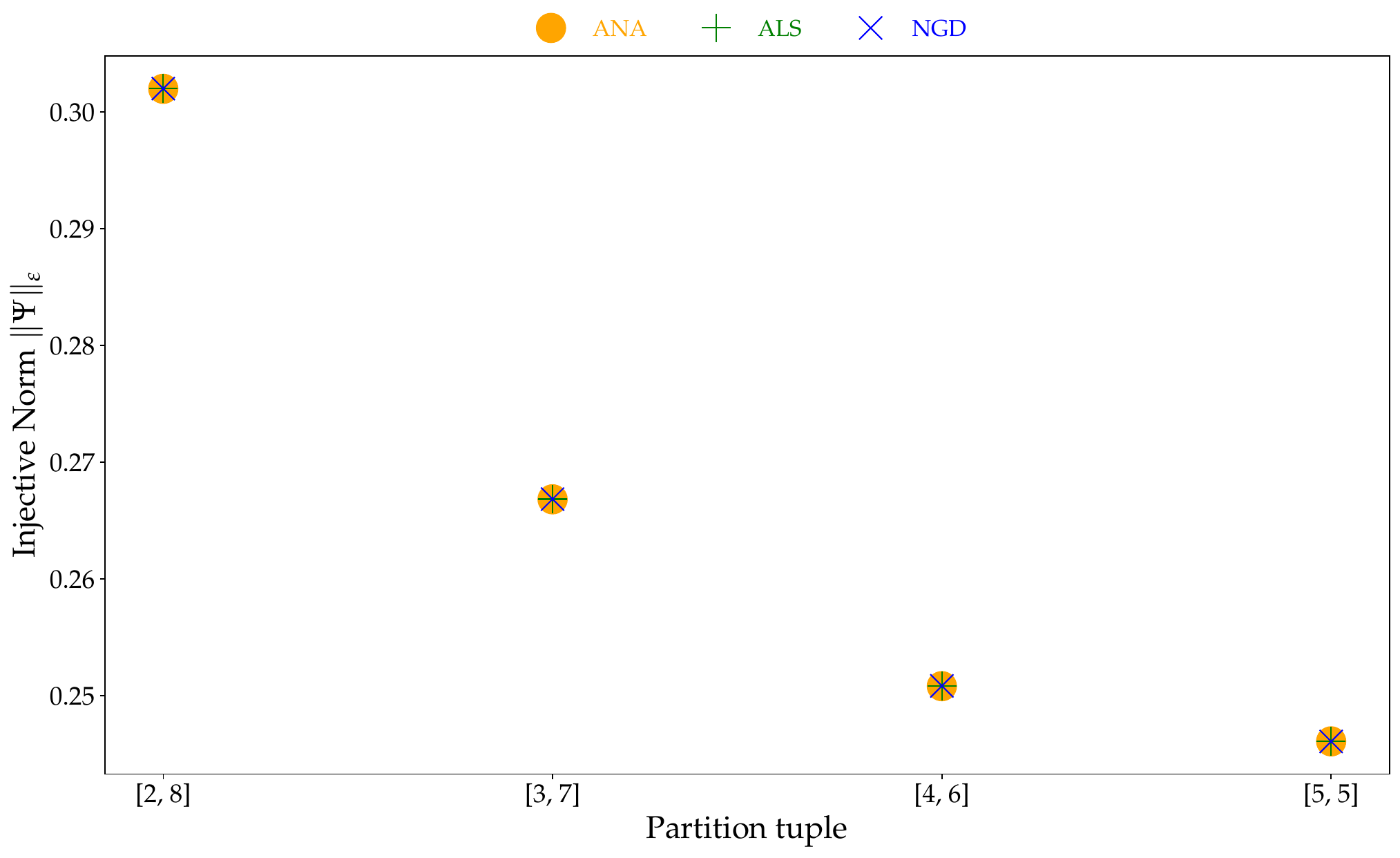}
    \label{fig:dicke_inj}
    \caption{We consider generalized Dicke states with $n=10$ and \(d=2\) as it holds physical relevance \cite{hume2009dicke}. Both the ALS and NGD algorithms perform equally well at approximating the value of the normalized injective norm \(\|\hat\Psi_D(10,2,\vec{k})\|_\varepsilon\), regardless of \(\vec{k}\).
    }
\end{figure}

\subsection{Antisymmetric basis states}
Antisymmetric states have found substantial application in quantum cryptography \cite{jex2003antisym} and in the study of fermionic systems \cite{zhu2010additivity}. An antisymmetric pure state \(\Psi_A \in \left(\mathbb{C}^d\right)^{\otimes n}\) is a normalized tensor such that every odd permutation of its tensor factors induces a sign change. The antisymmetric subspace of \(\left(\mathbb{C}^d\right)^{\otimes n}\) is the collection of all pure antisymmetric tensors and has a dimension \(\binom{d}{n}\), where \(d \geq n\). The antisymmetric basis states can be constructed using the permutations of an orthonormal family \(\{a_1,\ldots,a_n\}\) of \(\mathbb C^d\), as

\begin{equation}\label{eqn:anti_sym}
    \Psi_A(n, d) := \frac{1}{\sqrt{n!}} \sum_{\sigma \in S_n} \operatorname{sgn}(\sigma)a_{\sigma(1)}\otimes a_{\sigma(2)}\otimes\cdots\otimes a_{\sigma(n)},
\end{equation}

where \(S_n\) is the permutation group of \(n\) elements, \(\operatorname{sgn(\sigma)}\) is the sign of the permutation \(\sigma\) and \(\frac{1}{\sqrt{n!}}\) is the normalization factor. The analytical value for the GME of such states is known \cite{zhu2010additivity} and is given by
\begin{equation}
\GME\left(\Psi_A(n,d)\right) = \log (n!) .
\end{equation}
\begin{figure}[!h]
    \centering
    \includegraphics[width=0.9\textwidth]{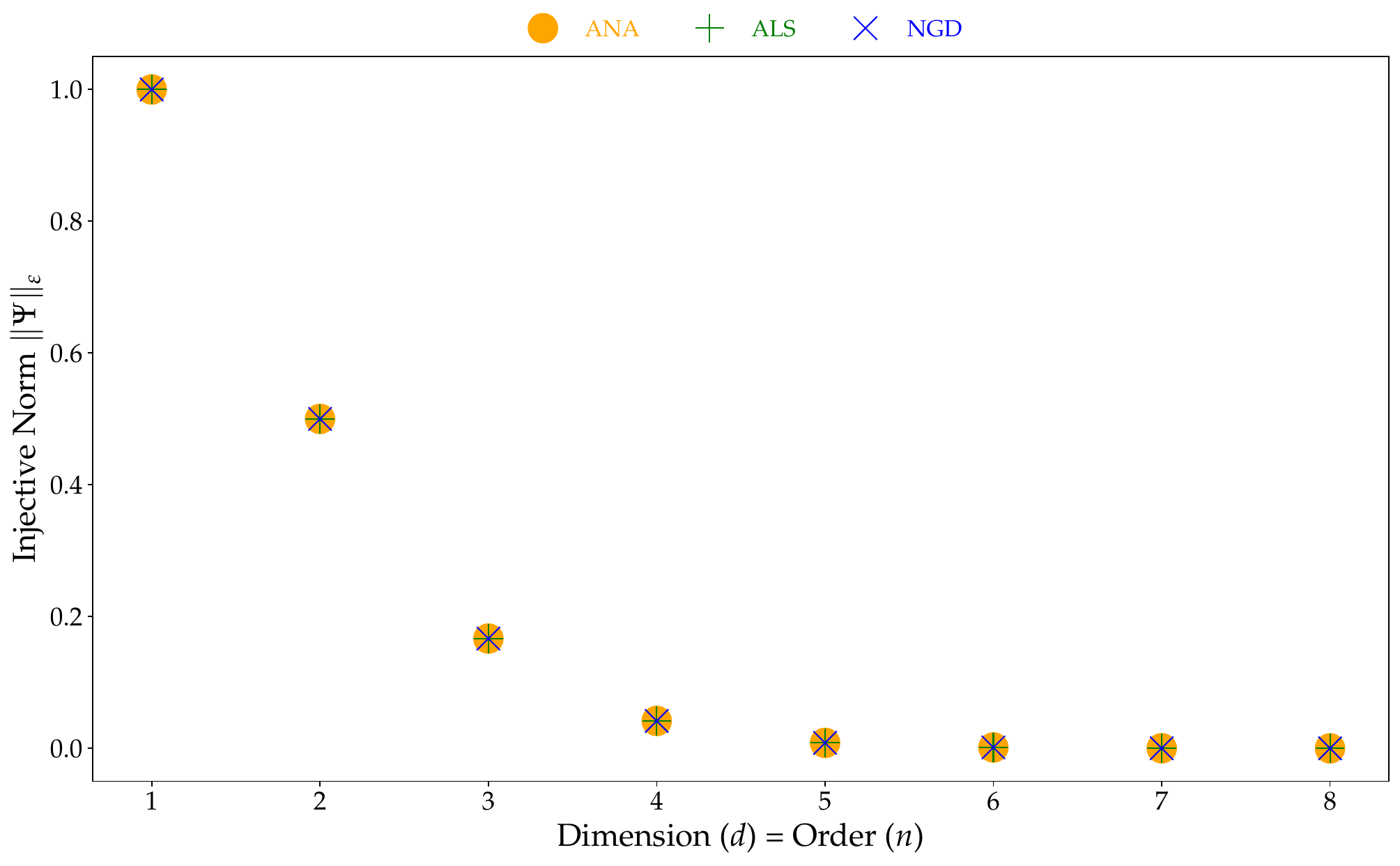}\vspace{-2mm}
    \caption{We consider antisymmetric basis states with \(d=n\) from \(1\) to \(8\). Both the ALS and NGD algorithms perform equally well at approximating the value of the normalized injective norm \(\|\hat\Psi_A(n,d)\|_\varepsilon\).\looseness-1}
    \label{fig:antisym_inj}
\end{figure}

When benchmarking ALS and NGD on deterministic states, such as generalized Dicke states and antisymmetric basis states, we observe that both algorithms always perform equally well. This is in accordance with our observations in Section \ref{sec:benchmark-real}, namely that ALS and NGD have similar performances on normalized inputs. We emphasize that here, we are interested in estimating the injective norm of specific tensors with fixed (in fact usually quite small) local dimension \(d\). So there is no problem of having to deal with normalized injective norms that would go to \(0\) as \(d\) grows, as opposed to the case of random tensors, where we were trying to grasp the asymptotic values in the large dimension or order regime. Hece, in these kinds of situations, our tests suggest that one can indifferently use ALS or NGD.\\ 

\newpage
\paragraph{Open-source commitment}
The algorithms that we develop, mainly NGD and its symmetrized version SGD, can be used to approximate the injective norm of any tensor (SGD for any symmetrized tensor), real or complex, regardless of its number of tensor factors and its local dimension (provided sufficient computational resources). We have published our code in Python with all the features used in this work as an open-source repository along with installation instructions \cite{code}. Further, we will continue maintaining the repository and integrating new features based on community feedback.

\end{document}